\documentclass[twocolumn,english,aps,prb,twocolumn,superscriptaddress,showpacs,a4paper]{revtex4-1}
\usepackage{lmodern}

\usepackage[T1]{fontenc}
\usepackage[latin9]{inputenc}
\usepackage[a4paper]{geometry}
\usepackage{color}
\usepackage{babel}
\usepackage{prettyref}
\usepackage{booktabs}
\usepackage{bm}
\usepackage{amsmath}
\usepackage{amssymb}
\usepackage{graphicx}
\usepackage{esint}
\usepackage{subscript}
\usepackage[unicode=true,pdfusetitle,
 bookmarks=true,bookmarksnumbered=false,bookmarksopen=false,
 breaklinks=false,pdfborder={0 0 0},backref=false,colorlinks=true]
 {hyperref}
\hypersetup{
 urlcolor=blue,linkcolor=blue,citecolor=blue}

\makeatletter

\providecommand{\tabularnewline}{\\}

\@ifundefined{textcolor}{}
{%
 \definecolor{BLACK}{gray}{0}
 \definecolor{WHITE}{gray}{1}
 \definecolor{RED}{rgb}{1,0,0}
 \definecolor{GREEN}{rgb}{0,1,0}
 \definecolor{BLUE}{rgb}{0,0,1}
 \definecolor{CYAN}{cmyk}{1,0,0,0}
 \definecolor{MAGENTA}{cmyk}{0,1,0,0}
 \definecolor{YELLOW}{cmyk}{0,0,1,0}
 }

\usepackage{upgreek}
\usepackage{textcomp}
\newcommand{\murm}{%
  \ifmmode
    \mathchoice
        {\hbox{\normalsize\textmu}}
        {\hbox{\normalsize\textmu}}
        {\hbox{\scriptsize\textmu}}
        {\hbox{\tiny\textmu}}%
  \else
    \textmu
  \fi
}
\newcommand{\muon}{\murm$^{\text{+}}$}
\newcommand{\muSR}{\muon SR}
\newcommand{\Fmu}{F\murm}
\newcommand{\FmuF}{F\murm F}
\newrefformat{tab}{Table~\ref{#1}}
\newrefformat{fig}{Fig.~\ref{#1}}
\newrefformat{eq}{Eq.~\textup{(\ref{#1})}}
\newrefformat{sec}{Sec.~\ref{#1}}
\newrefformat{sub}{Sec.~\ref{#1}}

\makeatother

\begin{document}

\title{Magnetic order in quasi--two-dimensional molecular magnets \\ investigated with muon-spin relaxation}

\author{A. J. Steele}

\author{T. Lancaster}

\author{S. J. Blundell}

\affiliation{Oxford University Department of Physics, Clarendon Laboratory, Parks
Road, Oxford. OX1 3PU, United Kingdom}

\author{P. J. Baker}

\author{F. L. Pratt}

\affiliation{ISIS Pulsed Neutron and Muon Source, STFC Rutherford Appleton Laboratory,
Harwell Science and Innovation Campus, Didcot, Oxfordshire. OX11 0QX,
United Kingdom}

\author{C. Baines}

\affiliation{Paul Scherrer Institut, Laboratory for Muon-Spin Spectroscopy, CH-5232
Villigen PSI, Switzerland.}

\author{M. M. Conner}

\author{H. I. Southerland}

\author{J. L. Manson}

\affiliation{Department of Chemistry and Biochemistry, Eastern Washington University,
Cheney. WA 99004, USA }

\author{J. A. Schlueter}

\affiliation{Materials Science Division, Argonne National Laboratory. Argonne
IL 60439, USA}

\date{\today}
\begin{abstract}
We present the results of a muon-spin relaxation (\muSR) investigation
into magnetic ordering in several families of layered quasi--two-dimensional
molecular antiferromagnets based on transition metal ions such as
$S=\frac{1}{2}$ Cu$^{\text{2+}}$ bridged with organic ligands such
as pyrazine. In many of these materials magnetic ordering is difficult
to detect with conventional magnetic probes. In contrast, \muSR\ allows
us to identify ordering temperatures and study the critical behavior
close to $T_{\mathrm{N}}$. Combining this with measurements of in-plane
magnetic exchange $J$ and predictions from quantum Monte Carlo simulations
we may assess the degree of isolation of the 2D layers through estimates
of the effective inter-layer exchange coupling and in-layer correlation
lengths at $T_{\mathrm{N}}$. We also identify the likely metal-ion
moment sizes and muon stopping sites in these materials, based on
probabilistic analysis of the magnetic structures and of muon--fluorine
dipole--dipole coupling in fluorinated materials.
\end{abstract}

\pacs{76.75.+i, 75.50.Xx, 75.10.Jm, 75.50.Ee}

\maketitle

\section{Introduction}

The $S=\frac{1}{2}$ two-dimensional square-lattice quantum Heisenberg
antiferromagnet (2DSLQHA) continues to be one of the most important
theoretical models in condensed matter physics~\cite{Manousakis1991-2DSLHAFM}.
Experimental realizations of the 2DSLQHA in crystals also contain
an interaction between planes, so that the relevant model describing
the coupling of electronic spins $\bm{S}_{i}$ gives rise to the Hamiltonian%
\footnote{Note that, in this model, the exchange energy in a bond between two
parallel spins is $2J$. The sums are therefore over unconstrained
values of $i$ and $j$. They include an implicit factor of $\frac{1}{2}$
to prevent double-counting, leading to the form in \prettyref{eq:2D-Hamiltonian}.%
} 
\begin{equation}
\mathcal{H}=J\sum_{\langle i,j\rangle_{xy}}\bm{S}_{i}\cdot\bm{S}_{j}+J_{\perp}\sum_{\langle i,j\rangle_{z}}\bm{S}_{i}\cdot\bm{S}_{j},\label{eq:2D-Hamiltonian}
\end{equation}
where $J$ ($J_{\perp}$) is the strength of the in- (inter-) plane
coupling and the first (second) summation is over neighbors parallel
(perpendicular) to the 2D $xy$-plane. Any 2D model ($J_{\perp}=0$)
with continuous symmetry will not show long-range magnetic order (LRO)
for $T>0$ due to a divergence of infrared fluctuations~\cite{Mermin1966-and-Wagner-no-1-2D-order,Berezinskii1971-Mermin-and-Wagner}.
However, layered systems approximating 2D models ($J_{\perp}\neq0$)
will inevitably enjoy some degree of interlayer coupling and this
will lead to magnetic order, albeit at a reduced temperature due to
the influence of quantum fluctuations. Quantum fluctuations are also
predicted to reduce the value of the magnetic moment in the ground
state of the 2DSLQHA to around 60\% of its classical value \cite{Manousakis1991-2DSLHAFM},
and this reduction is often seen in the ordered moments of real materials.
In layered materials that approximate the 2DSLQHA, the measurement
of the antiferromagnetic ordering temperature $T_{\mathrm{N}}$ is
often problematic due not only to this reduction of the magnetic moment,
but also to short-range correlations that build up in the quasi-2D
layers above $T_{\mathrm{N}}$. These correlations lead to a reduction
in the size of the entropy change that accompanies the phase transition,
reducing the size of the anomaly in the measured specific heat~\cite{sengupta2003-bulkmeasuresbad}.
We have shown in a number of previous cases that muon-spin relaxation
(\muSR) measurements do not suffer from these effects and therefore
represent an effective method for detecting magnetic order in complex
anisotropic systems~\cite{Blundell2007-muSR-lowD-magnets-review,goddard2008-2DHMexchange,lancaster2007-CuPz2ClO42,lancaster2006-CuPzN}.

The rich chemistry of molecular materials allows for the design and
synthesis of a wide variety of highly-tunable magnetic model systems
\cite{Blundell2004-molmagreview}. Magnetic centers, exchange paths
and the surrounding molecular groups can all be systematically modified,
allowing investigation of their effects on magnetic behavior. In particular,
the existence of different exchange paths along different spatial
directions can result in quasi--low-dimensional magnetic behavior
(i.e.\ systems with magnetic interactions constrained to act in a
two-dimensional plane or along a one-dimensional chain). Such systems
have the potential to better approximate low-dimensional models than
many traditional inorganic materials. In addition, these molecular
materials can have exchange energy scales of order $J/k_{\mathrm{B}}\sim10\mathrm{\; K}$
which are accessible with typical laboratory magnetic fields \cite{goddard2008-2DHMexchange}
of $B\sim10\mathrm{\; T}$ allowing an additional avenue for their
experimental study. This contrasts with typical inorganic low-dimensional
systems where the exchange is found to be $J/k_{\mathrm{B}}\sim1000\mathrm{\; K}$
and fields of $B\sim1000\mathrm{\; T}$ would be needed to significantly
perturb the spin system.

It has also been shown~\cite{Xiao2009-XY-q2DHA,Sengupta2009-nonmonotonic-TN-B-CuHF2pyz2BF4}
that a small $XY$-like anisotropy exists in some molecular materials.
Although $J_{\perp}$ is the decisive energy scale for the magnetic
ordering, this anisotropy has been shown to have an influence on the
ordering temperature~\cite{Xiao2009-XY-q2DHA} and determines the
shape of the low-field $B$--$T$ phase diagram of these systems~\cite{Sengupta2009-nonmonotonic-TN-B-CuHF2pyz2BF4}.

\begin{figure}
\includegraphics{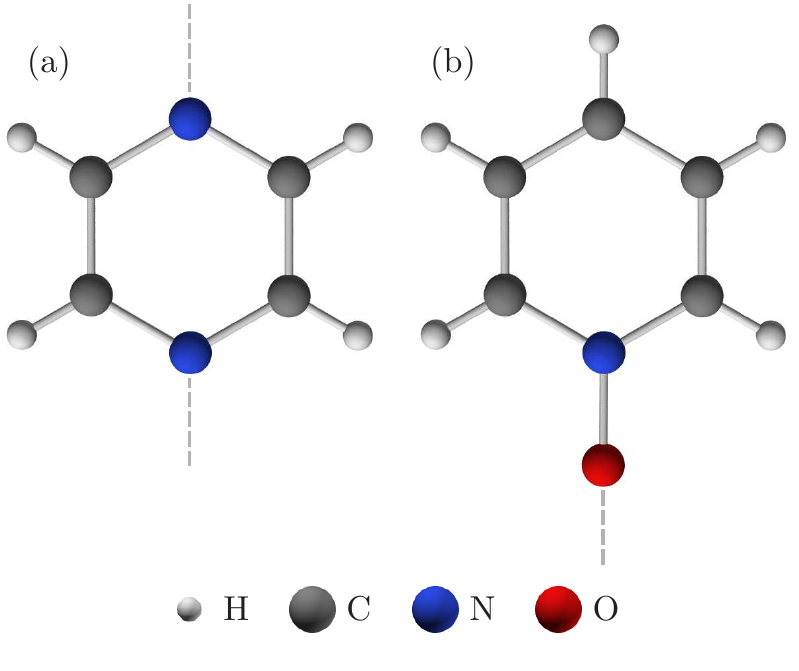}

\caption{Bridging ligands used in the compounds described in this paper: (a)
pyrazine (N$_{\text{2}}$C$_{\text{4}}$H$_{\text{4}}$, abbreviated
pyz); and (b) (ii) pyridine-\emph{N}-oxide (C$_{\text{5}}$H$_{\text{5}}$NO,
abbreviated pyo). Dashed lines indicate where the ligands bond to
other parts of molecular structures.\label{fig:ligands}}
\end{figure}
Several classes of molecular magnetic material closely approximate
the 2DSLQHA model and in this paper we report the results of \muSR\
measurements performed on several such materials. These systems are
self-assembled coordination polymers, based around paramagnetic ions
such as Cu$^{\text{2+}}$, linked by neutral bridging ligands and
coordinating anion molecules. Our materials are based on combinations
of three different ligands: (i) pyrazine (N$_{\text{2}}$C$_{\text{4}}$H$_{\text{4}}$,
abbreviated pyz) and (ii) pyridine-\emph{N}-oxide (C$_{\text{5}}$H$_{\text{5}}$NO,
abbreviated pyo), both of which are planar rings; and (iii) the linear
bifluoride ion {[}(HF$_{\text{2}}$)$^{-}${]}, which is bound by
strong hydrogen bonds F$^{\text{...}}$H$^{\text{...}}$F. The pyz
and pyo ligands are shown in \prettyref{fig:ligands}. Specifically,
we investigate the molecular system {[}\emph{M}(HF$_{\text{2}}$)(pyz)$_{\text{2}}${]}\emph{X},
where \emph{M}$^{\text{2+}}$~= Cu$^{\text{2+}}$ is the transition
metal cation and \emph{X}$^{-}$ is one of various anions (e.g.\ BF$_{\text{4}}^{-}$,
ClO$_{\text{4}}^{-}$, PF$_{\text{6}}^{-}$ etc.). We also report
the results of our measurements on other quasi-2D systems. First,
{[}Cu(pyz)$_{\text{2}}$(pyo)$_{\text{2}}${]}\emph{Y}$_{\text{2}}$,
with \emph{Y}$^{-}$~= BF$_{\text{4}}^{-}$ or PF$_{\text{6}}^{-}$,
in which pyo ligands bridge Cu(pyz)$_{\text{2}}$ planes. Then, the
quasi-2D non-polymeric compounds {[}Cu(pyo)$_{\text{6}}${]}\emph{Z}$_{\text{2}}$,
where \emph{Z}$^{-}$=BF$_{\text{4}}^{-}$, ClO$_{\text{3}}^{-}$
or PF$_{\text{6}}^{-}$ are examined. We also investigate materials
in which either Ni$^{\text{2+}}$ ($S=1$) or Ag$^{\text{2+}}$ ($S=\frac{1}{2}$)
form the magnetic species in the quasi-2D planes rather than Cu$^{\text{2+}}$.

This paper is structured as follows. In \prettyref{sec:Experimental-details}
we outline the \muSR\ technique and describe our experimental methods.
The {[}Cu(HF$_{\text{2}}$)(pyz)$_{\text{2}}${]}\emph{X} family of
materials is then discussed in \prettyref{sec:[Cu(HF2)(pyz)2]X},
where muon data are used to determine $T_{\mathrm{N}}$ and critical
parameters. Muon--fluorine dipole--dipole oscillations in the paramagnetic
regime are found for these materials which we use, in conjunction
with dipole field simulations, to investigate possible muon sites
and constrain the copper moment. In \prettyref{sec:[Cu(pyz)2(pyo)2]X2}
we explore the related 2D system {[}Cu(pyz)$_{\text{2}}$(pyo)$_{\text{2}}${]}\emph{Y}$_{\text{2}}$.
\prettyref{sec:[Cu(pyo)6]X2} details measurements of {[}Cu(pyo)$_{\text{6}}${]}\emph{Z}$_{\text{2}}$.
\prettyref{sec:Agpyz2S2O8} examines a highly two-dimensional silver-based
molecular material, Ag(pyz)$_{\text{2}}$(S$_{\text{2}}$O$_{\text{8}}$).
Finally, data from the {[}Ni(HF$_{\text{2}}$)(pyz)$_{\text{2}}${]}\emph{X}
(\emph{X}$^{-}$\emph{~=} PF$_{\text{6}}^{-}$, SbF$_{\text{6}}^{-}$)
family of molecular magnets is presented in \prettyref{sec:dog}.

\section{Experimental details\label{sec:Experimental-details}}

Zero-field (ZF) \muSR\ measurements were made on powder samples of
the materials at the ISIS facility, Rutherford Appleton Laboratory,
UK using the MuSR and EMU instruments and the Swiss Muon Source (S\murm S),
Paul Scherrer Institut, Switzerland using the General-Purpose Surface-Muon
(GPS) instrument and Low-Temperature Facility (LTF). For measurements
at temperatures $T\geq1.8\mathrm{\; K}$ powder samples were packed
in a $25\mathrm{\;\murm m}$ Ag foil packet and mounted on a Ag backing
plate. For measurements at $T<1.8\mathrm{\; K}$ the samples were
mounted directly on an Ag plate and covered with a $12.5\mathrm{\;\murm m}$
Ag foil mask. Ag is used since it has only a small nuclear magnetic
moment, and so minimizes the background depolarization of the muon
spin ensemble.

In a \muSR\ experiment~\cite{blundell1999-mureview}, spin-polarized
positive muons are stopped in a target sample. The positive muons
are attracted to areas of negative charge density and often stop at
interstitial positions. The observed property of the experiment is
the time evolution of the muon-spin polarization, the behaviour of
which depends on the local magnetic field at the muon site. Each muon
decays with an average lifetime of $2.2\mathrm{\;\murm s}$ into two
neutrinos and a positron, the latter particle being emitted preferentially
along the instantaneous direction of the muon spin. Recording the
time dependence of the positron emission directions therefore allows
the determination of the spin polarization of the ensemble of muons.
In our experiments, positrons are detected by detectors placed forward~($\mathrm{F}$)
and backward~($\mathrm{B}$) of the initial muon polarization direction.
Histograms $N_{\mathrm{F}}(t)$ and $N_{\mathrm{B}}(t)$ record the
number of positrons detected in the two detectors as a function of
time following the muon implantation. The quantity of interest is
the decay positron asymmetry function, defined as 
\begin{equation}
A(t)=\frac{N_{\mathrm{F}}(t)-\alpha N_{\mathrm{B}}(t)}{N_{\mathrm{F}}(t)+\alpha N_{\mathrm{B}}(t)}\,,\label{eq:asymmetry-general}
\end{equation}
 where $\alpha$ is an experimental calibration constant. The asymmetry,
$A(t)$, is proportional to the spin polarization of the muon ensemble.

\begin{figure}
\includegraphics{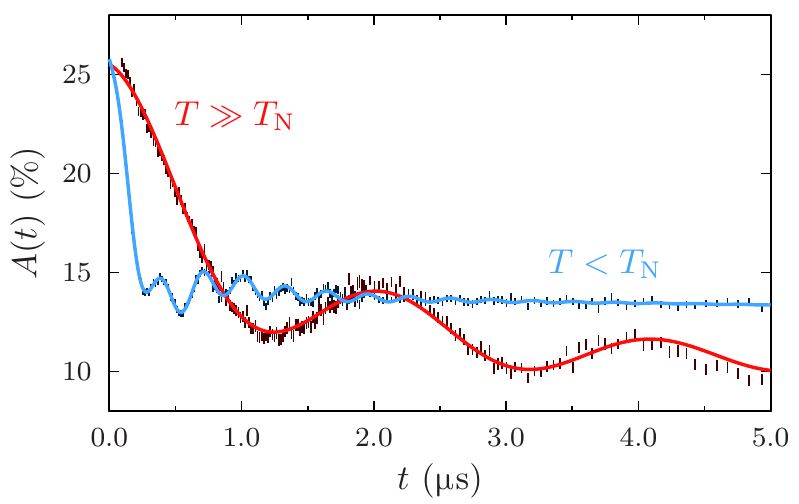} \caption{Example $A(t)$ spectra with fits above ($T=25$~K) and below ($T=0.35$~K)
the magnetic transition temperature $T_{\mathrm{N}}=1.4$~K for {[}Cu(HF$_{\text{2}}$)(pyz)$_{\text{2}}${]}BF$_{\text{4}}$.
Note the approximately equal initial asymmetry, and the oscillations
and `$\frac{1}{3}$-tail' observed in the ordered phase. The slow
oscillation observed for $T>T_{\mathrm{N}}$ is due to muon--fluorine
dipole--dipole oscillations. \label{fig:eg-A(t)-spectra}}
\end{figure}

A muon spin will precess around the local magnetic fields at its stopping
site at a frequency $\nu=\gamma_{\murm}B/2\pi$, where the muon gyromagnetic
ratio $\gamma_{\murm}=2\pi\times135.5\mathrm{\; MHz\, T^{-1}}$. %When averaged over the ensemble of muons, 
%this
%gives rise to two broad classes of behaviour: (i) a monotonic relaxation of $A(t)$
%to a state of zero muon-spin polarization
%often indicates that the muon spins are being depolarised due to a
%combination of a wide distribution internal magnetic fields and dynamic fluctuations; 
In the presence of LRO in a material we often measure oscillations
in $A(t)$. These result from a significant number of muons stopping
at sites with a similar internal field, giving rise to a coherent
precession of the ensemble of muon spins. Since the spins precess
in local magnetic fields directed perpendicularly to the spin polarization
direction, we would expect that, for a powder sample with a static
magnetic field distribution, $\frac{2}{3}$ of the total spin components
should precess and the remaining $\frac{1}{3}$ should be non-relaxing.
The non-relaxing third of muon-spin components give rise to the so-called
$\frac{1}{3}$-tail in $A(t)$, whose presence therefore provides
additional evidence for a static field distribution in powder sample.
Taken together, these effects provide an unambiguous method for sensitively
identifying a transition to LRO. An example of typical spectra above
and below the magnetic ordering temperature is shown for {[}Cu(HF$_{\text{2}}$)(pyz)$_{\text{2}}${]}BF$_{\text{4}}$
in \prettyref{fig:eg-A(t)-spectra}; the oscillations above $T_{\mathrm{N}}$
are characteristic of a quantum-entangled \Fmu\  state (see \prettyref{sub:[Cu(HF2)(pyz)2]X-paramagnetic}).

\section{$\text{[Cu(HF}_{\text{2}}\text{)(pyz)}_{\text{2}}\text{]}\text{\emph{X}}$\label{sec:[Cu(HF2)(pyz)2]X}}

\begin{figure}
\includegraphics{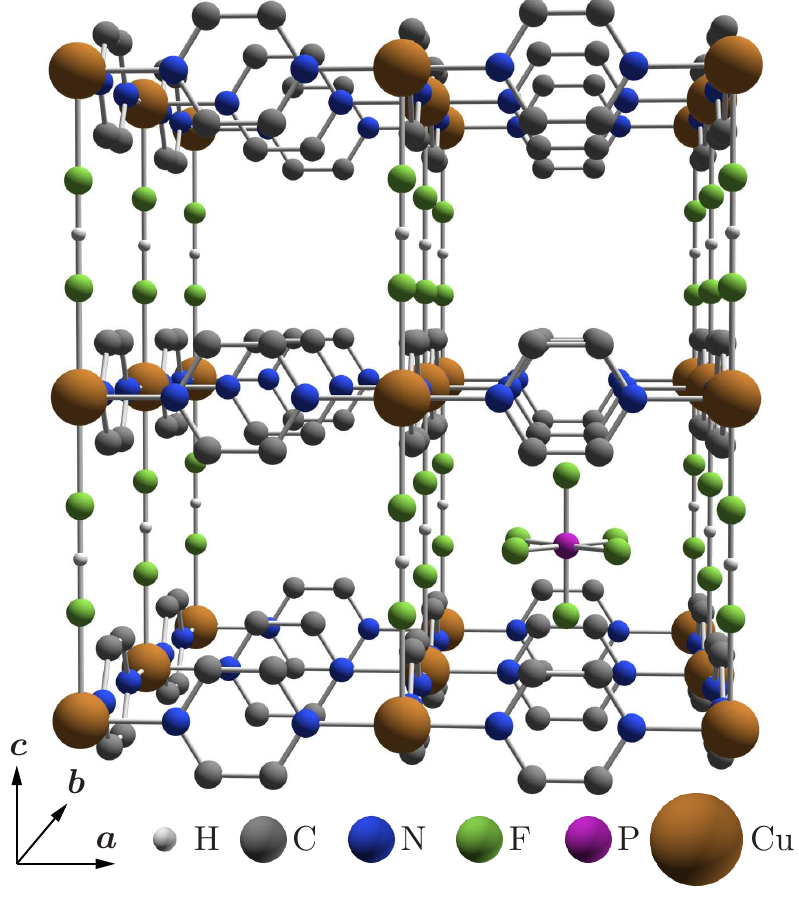}
\caption{The structure of {[}Cu(HF$_{\text{2}}$)(pyz)$_{\text{2}}${]}PF$_{\text{6}}$,
as an example of the {[}\emph{M}(HF$_{\text{2}}$)(pyz)$_{\text{2}}${]}\emph{X}
series. Copper ions are joined in a 2D square lattice by pyrazine
ligands to form Cu(pyz)$_{\text{2}}^{\text{2+}}$ sheets; the 2D layers
are joined in the third dimension by HF$_{\text{2}}^{-}$ groups,
making a pseudocubic 3D structure; and this structure is stabilised
by a PF$_{\text{6}}^{-}$ anion at the center of each cubic pore.
For clarity, hydrogen atoms attached to pyrazine rings have been omitted,
and only one PF$_{\text{6}}^{-}$ anion is shown.\label{fig:M(HF2)(pyz)2X-structure}}
\end{figure}

The synthesis of the {[}\emph{M}(HF$_{\text{2}}$)(pyz)$_{\text{2}}${]}\emph{X}
system~\cite{manson2006-CuHF2pyz2BF4-chemcomm,goddard2008-2DHMexchange,manson2009-HFsynthons-JACScover}
represented the first example of the use of a bifluoride building
block to make a three-dimensional coordination polymer. This class
of materials possesses a highly stable structure due to the exceptional
strength of the bifluoride hydrogen bonds. The structure of the {[}\emph{M}(HF$_{\text{2}}$)(pyz)$_{\text{2}}${]}\emph{X}
system~\cite{manson2006-CuHF2pyz2BF4-chemcomm,manson2009-HFsynthons-JACScover}
comprises infinite 2D {[}\emph{M}(pyz)$_{\text{2}}${]}$^{\text{2+}}$
sheets which lie in the $ab$ plane, with bifluoride ions (HF$_{\text{2}}$)$^{-}$
above and below the metal ions, acting as bridges between the planes
to form a pseudocubic network. The \emph{X}$^{-}$ anions occupy the
body-center positions within each cubic pore. An example structure,
for {[}Cu(HF$_{\text{2}}$)(pyz)$_{\text{2}}${]}PF$_{\text{6}}$,
is shown in Fig. \ref{fig:M(HF2)(pyz)2X-structure}. Samples are produced
in polycrystalline form via aqueous chemical reactions between \emph{MX}$_{\text{2}}$
salts and stoichiometric amounts of ligands. Preparation details for
the compounds are reported in Refs.~\onlinecite{manson2006-CuHF2pyz2BF4-chemcomm,manson2009-HFsynthons-JACScover,Manson2011-NiHF2pyz2X}.

In this section we consider those materials where the \emph{M} cations
are Cu$^{\text{2+}}$ 3d$^{\text{9}}$ $S=\frac{1}{2}$ centers. It
is thought that the magnetic behavior of these material results from
the $3d_{x^{2}-y^{2}}$ orbital of the Cu at the center of each CuN$_{\text{4}}$F$_{\text{2}}$
octahedron lying in the CuN$_{\text{4}}$ plane so that the spin exchange
interactions between neighboring Cu$^{\text{2+}}$ ions occur through
the $s$-bonded pyz ligands\cite{manson2006-CuHF2pyz2BF4-chemcomm}.
The interplane exchange through the HF$_{\text{2}}$ bridges connecting
two Cu$^{\text{2+}}$ ions should be very weak as these bridges lie
on the 4-fold rotational axis of the Cu $3d_{x^{2}-y^{2}}$ magnetic
orbital, resulting in limited overlap with the fluorine $p_{z}$ orbitals.
Therefore to a first approximation, the magnetic properties of {[}\emph{M}(HF$_{\text{2}}$)(pyz)$_{\text{2}}${]}\emph{X}
can be described in terms of a 2D square lattice.

Measurements for \emph{X}$^{-}$~= BF$_{\text{4}}^{-}$, ClO$_{\text{4}}^{-}$
and SbF$_{\text{6}}^{-}$ were made using the MuSR spectrometer at
ISIS, whilst PF$_{\text{6}}$, AsF$_{\text{6}}$, NbF$_{\text{6}}$
and TaF$_{\text{6}}$ were measured using GPS at PSI.

\subsection{Long-range magnetic order\label{sub:[Cu(HF2)(pyz)2]X-LRO}}

\begin{figure*}
\includegraphics{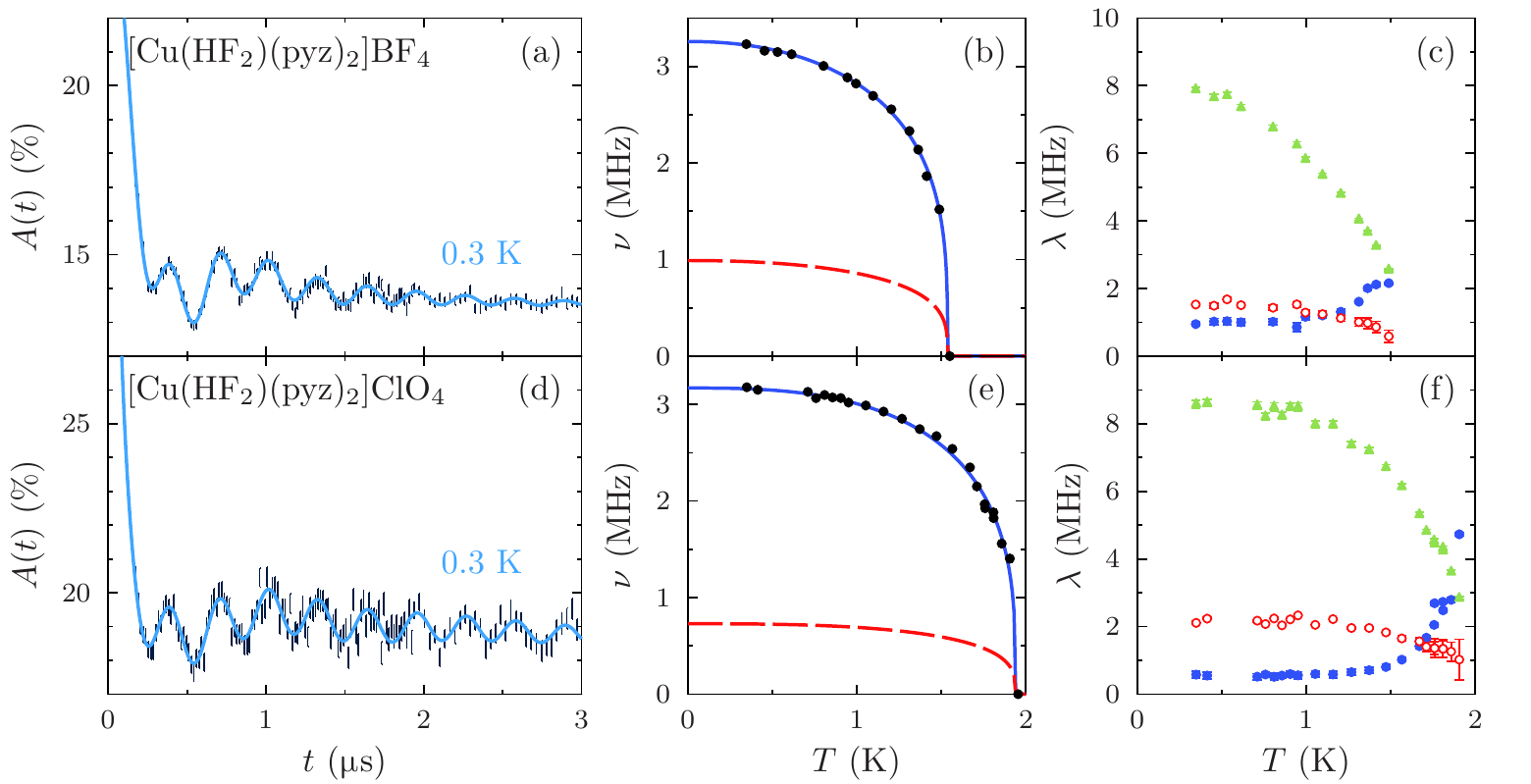}

\caption{Data and the results of fits to \prettyref{eq:M(HF2)(pyz)2X-A(t)-fit}
for {[}Cu(HF$_{\text{2}}$)(pyz)$_{\text{2}}${]}\emph{X} magnets
with tetrahedral anions \emph{X}$^{-}$. From left to right: (a) and
(d) show sample asymmetry spectra $A(t)$ for $T<T_{\mathrm{N}}$
along with a fit to \prettyref{eq:M(HF2)(pyz)2X-A(t)-fit}; (b) and
(e) show frequencies as a function of temperature {[}no data points
are shown for the second line because this frequency $\nu_{2}$ was
held in fixed proportion to the first, $\nu_{1}$ (see text){]}; and
(c) and (f) show relaxation rates $\lambda_{i}$ as a function of
temperature. In the $\nu(T)$ plot, error bars are included on the
points but in most cases they are smaller than the marker being used.
The solid line representing $\nu_{1}$ in (b) and (c) corresponds
to the filled circles in the third column of graphs {[}(c) and (f){]}
for that component's relaxation, $\lambda_{1}$, whilst the dashed
line and unfilled circles correspond to $\nu_{2}$ and $\lambda_{2}$,
respectively. The filled triangles correspond to the fast relaxation
$\lambda_{3}$. \label{fig:Cu(HF2)(pyz)2X-tetrahedral-magnetism-graphs}}
\end{figure*}

\begin{figure*}
\includegraphics{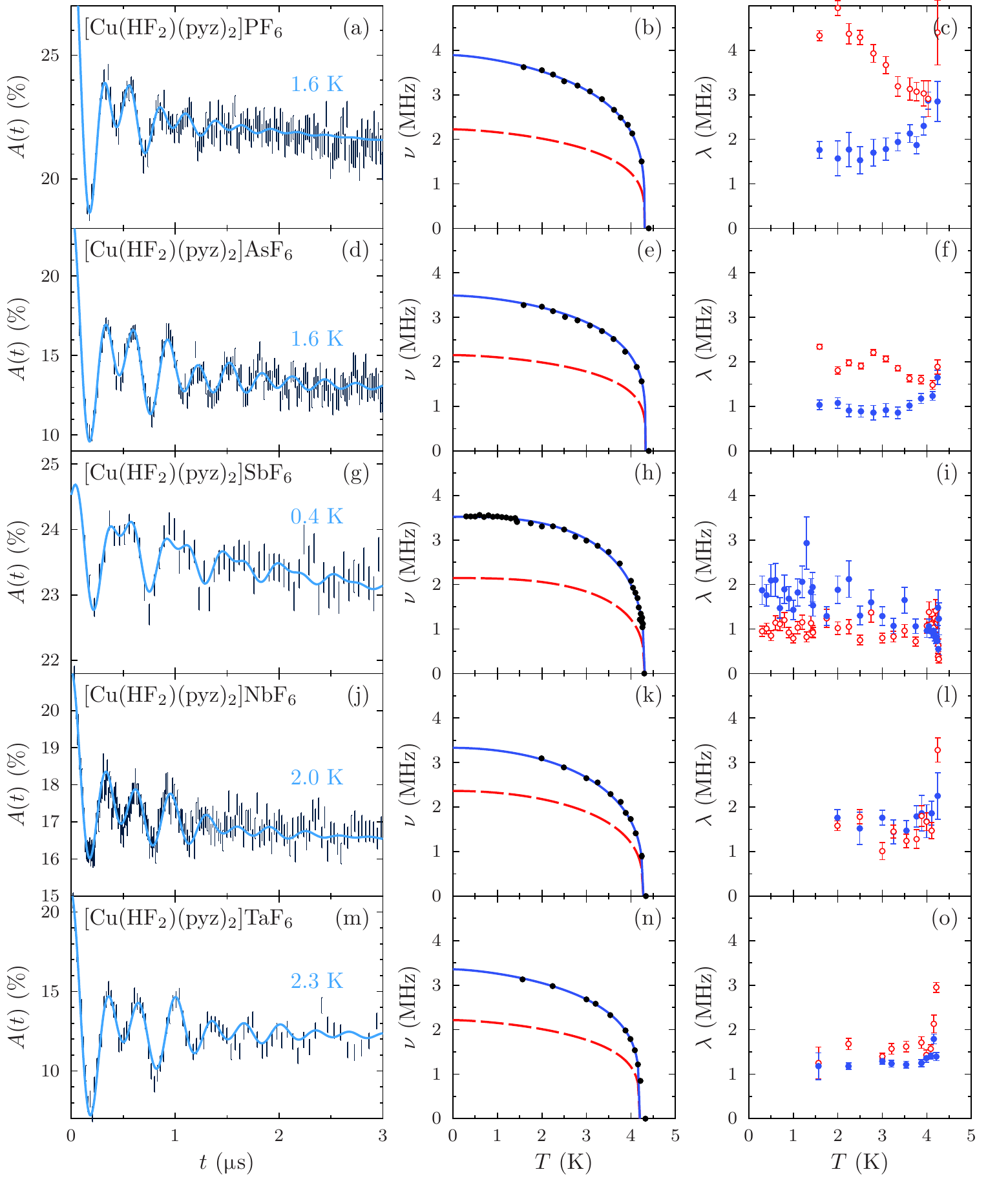}

\caption{Example data and fits for {[}Cu(HF$\text{\ensuremath{_{2}}}$)(pyz)$_{\text{2}}${]}\emph{X}
magnets with octahedral anions \emph{X}$^{-}$. From left to right:
(a), (d), (g), (j) and (m) show sample asymmetry spectra $A(t)$ for
$T<T_{\mathrm{N}}$ along with a fit to \prettyref{eq:M(HF2)(pyz)2X-A(t)-fit};
(b), (e), (h), (k) and (n) show frequencies as a function of temperature
{[}no data points are shown for the second line because this frequency
$\nu_{2}$ was held in fixed proportion to the first, $\nu_{1}$ (see
text){]}; and (c), (f), (i), (l) and (o) show relaxation rates $\lambda_{i}$
as a function of temperature. In the $\nu(T)$ plot, error bars are
included on the points but in most cases they are smaller than the
marker being used. The solid line representing $\nu_{1}$ in (b),
(e), (h), (k) and (n) corresponds to the filled circles in the third
column of graphs {[}(c), (f), (i), (l) and (o){]} for that component's
relaxation, $\lambda_{1}$, whilst the dashed line and unfilled circles
correspond to $\nu_{2}$ and $\lambda_{2}$, respectively.\label{fig:Cu(HF2)(pyz)2X-octahedral-magnetism-graphs}}
\end{figure*}

The main result of our measurements on these systems is that below
a critical temperature $T_{\mathrm{N}}$, oscillations in the asymmetry
spectra $A(t)$ are observed at two distinct frequencies, for all
materials in the series. This shows unambiguously that each of these
materials undergoes a transition to a state of LRO. Example asymmetry
spectra are shown in the left-hand column of \prettyref{fig:Cu(HF2)(pyz)2X-tetrahedral-magnetism-graphs}
and \prettyref{fig:Cu(HF2)(pyz)2X-octahedral-magnetism-graphs}. They
were found to be best fitted with a relaxation function 
\begin{eqnarray}
A(t)= & A_{0} & \left[p_{1}\mathrm{e}^{-\lambda_{1}t}\cos(2\pi\nu_{1}t+\phi_{1})\right.\nonumber \\
 &  & \left.+p_{2}\mathrm{e}^{-\lambda_{2}t}\cos(2\pi P_{2}\nu_{1}t+\phi_{2})+p_{3}\mathrm{e}^{-\lambda_{3}t}\right]\nonumber \\
 &  & +A_{\mathrm{bg}}\mathrm{e}^{-\lambda_{\mathrm{bg}}t},\label{eq:M(HF2)(pyz)2X-A(t)-fit}
\end{eqnarray}
where $A_{0}$ represents the contribution from those muons which
stop inside the sample and $A_{\mathrm{bg}}$ accounts for a relaxing
background signal due to those muons that stop in the silver sample
holder or cryostat tails, or with their spin parallel to the local
field. Of those muons which stop in the sample, $p_{1}$ indicates
the weighting of the component in an oscillating state with frequency
$\nu_{1}$; $p_{2}$ is the weighting of a lower-frequency oscillating
state with frequency $\nu_{2}$; and $p_{3}$ represents the weighting
of a component with a large relaxation rate $\lambda_{3}$. All parameters
were initially left free to vary. The second frequency was found to
vary with temperature in fixed proportion to $\nu_{1}$ via $\nu_{2}=P_{2}\nu_{1}$
for each material. The parameter $P_{2}$ was identified by fitting
the lowest-temperature $A(t)$ spectra where \prettyref{eq:M(HF2)(pyz)2X-A(t)-fit}
would be expected to most accurately describe the data, and subsequently
held fixed during the fitting procedure. Phase factors $\phi_{i}$
were also found to be necessary in some cases to obtain a reliable
fit. The parameters resulting from these fits are listed in \prettyref{tab:Cu(HF2)(pyz)2X-magnetism-parameters},
and data with fits are shown in Figs.~\ref{fig:Cu(HF2)(pyz)2X-tetrahedral-magnetism-graphs}
and \ref{fig:Cu(HF2)(pyz)2X-octahedral-magnetism-graphs}. We also
note here that the discontinuous nature of the change in all fitted
parameters and the form of the spectra at $T_{\mathrm{N}}$ strongly
suggest that these materials are magnetically ordered throughout their
bulk.

\begin{table*}
\begin{tabular}{cccccccccccccc}
\toprule 
\emph{X}  & $\nu_{1}\mathrm{\:(MHz)}$  & $\nu_{2}\mathrm{\:(MHz)}$  & $\lambda_{3}\mathrm{\;(MHz)}$  & $p_{1}$  & $p_{2}$  & $p_{3}$  & $\phi_{1}\;(^{\circ})$ & $\phi_{2}\;(^{\circ})$ & $T_{\mathrm{N}}\mathrm{\:(K)}$  & $\beta$  & $\alpha$  & $J/k_{\mathrm{B}}\mathrm{\:(K)}$  & $|J_{\perp}/J|$\tabularnewline
\midrule 
BF$_{\text{4}}$  & $3.30(6)$  & $0.95(3)$  & $8$  & $15$  & $15$  & $70$  & $-57(1)$  & $26(2)$  & $1.54(2)$  & $0.18(4)$  & $1.6(7)$  & $6.3$  & $9\times10^{-4}$\tabularnewline
ClO$_{\text{4}}$  & $3.2(1)$  & $0.64(1)$  & $8$  & $2$  & $2$  & $96$  & $-94(1)$  & $104(4)$  & $1.91(1)$  & $0.25(2)$  & $2.6(3)$  & $7.3$  & $2\times10^{-3}$\tabularnewline
PF$_{\text{6}}$  & $3.89(6)$  & $2.23(5)$  & $25$  & $30$  & $30$  & $40$  & $-53(3)$  & $37(6)$  & $4.37(2)$  & $0.26(2)$  & $1.5(3)$  & $12.4$  & $1\times10^{-2}$\tabularnewline
AsF$_{\text{6}}$  & $3.49(9)$  & $2.15(2)$  & $40$  & $25$  & $25$  & $50$  & $-14(2)$  & $14(3)$  & $4.32(3)$  & $0.23(3)$  & $1.6(5)$  & $12.8$  & $1\times10^{-2}$\tabularnewline
SbF$_{\text{6}}$  & $3.51(2)$  & $2.14(2)$  & -  & $50$  & $50$  & -  & $-46(3)$  & $-20(2)$  & $4.29(1)$  & $0.34(2)$  & $2.8(3)$  & $13.3$  & $9\times10^{-3}$\tabularnewline
NbF$_{\text{6}}$  & $3.33(7)$  & $2.36(5)$  & $1$  & $40$  & $20$  & $40$  & $0$ & $-30(5)$  & $4.28(1)$  & $0.33(3)$  & $2.0(4)$  & -  & -\tabularnewline
TaF$_{\text{6}}$  & $3.33(9)$  & $2.21(5)$  & -  & $50$  & $50$  & -  & $0$ & $0$ & $4.22(1)$  & $0.25(1)$  & $1.5(3)$  & -  & -\tabularnewline
\bottomrule
\end{tabular}\caption{Fitted parameters for molecular magnets in the {[}Cu(HF$_{\text{2}}$)(pyz)$_{\text{2}}${]}\emph{X}
family. The first parameters shown relate to fits to \prettyref{eq:M(HF2)(pyz)2X-A(t)-fit},
which allow us to derive frequencies at $T=0$, $\nu_{i}$; probabilities
of stopping in the various classes of stopping site, $p_{i}$, in
percent; and phases associated with fitting the oscillating components
$\phi_{i}$. Then, the temperature dependence of $\nu_{i}$ is fitted
with \prettyref{eq:nuT1-T-over-TN-alpha-beta}, extracting values
for the Néel temperature, $T_{\mathrm{N}}$, critical exponent $\beta$
and parameter $\alpha$. Finally, the quoted $J/k_{\mathrm{B}}$ is
obtained from pulsed-field experiments~\cite{goddard2008-2DHMexchange},
and the ratio of inter- to in-plane coupling, $J_{\perp}/J$, is obtained
by combining $T_{\mathrm{N}}$ and $J$ with formulae extracted from
quantum Monte Carlo simulations (see \prettyref{sub:exchange-anisotropy},
and Ref.~\onlinecite{goddard2008-2DHMexchange}). Dashes in the $\lambda_{3}$
column for the SbF$_{\text{6}}$ and TaF$_{\text{6}}$ compounds indicate
that no fast-relaxing component was used to fit those data. Dashes
in the $J/k_{\mathrm{B}}$ and $J_{\perp}/J$ columns for NbF$_{\text{6}}$
and TaF$_{\text{6}}$ indicates a lack of pulsed-field data for these
materials. \label{tab:Cu(HF2)(pyz)2X-magnetism-parameters}}
\end{table*}

The frequencies and relaxation rates as a function of temperature
extracted from these fits are shown in the central column of Figs.~\ref{fig:Cu(HF2)(pyz)2X-tetrahedral-magnetism-graphs}
and \ref{fig:Cu(HF2)(pyz)2X-octahedral-magnetism-graphs}. The muon
precession frequency, which is proportional to the internal field
in the material, can be considered an effective order parameter for
the system. Consequently, fitting extracted frequencies as a function
of temperature to the phenomenological function 
\begin{equation}
\nu(T)=\nu(0)\left[1-\left(\frac{T}{T_{\mathrm{N}}}\right)^{\alpha}\right]^{\beta},\label{eq:nuT1-T-over-TN-alpha-beta}
\end{equation}
allows an estimate of the critical temperature and the exponent $\beta$
to be extracted. Our results fit well with a previous observation~\cite{goddard2008-2DHMexchange}
that the compounds divide naturally into two classes: those with tetrahedral
anions \emph{X}$^{-}$~=\emph{ }BF$_{\text{4}}^{-}$, ClO$_{\text{4}}^{-}$
and those with octahedral anions \emph{X}$^{-}$~= \emph{A}F$_{\text{6}}^{-}$.
The tetrahedral compounds have lower transition temperatures $T_{\mathrm{N}}\lesssim2\mathrm{\; K}$,
as compared to the octahedral compounds' $T_{\mathrm{N}}\gtrsim4\;\mathrm{K}$;
and the tetrahedral compounds also display slightly lower oscillation
frequencies than their octahedral counterparts~\cite{goddard2008-2DHMexchange}.

This difference has been explained~ in terms of differences in the
crystal structure between the two sets of compounds. Firstly, the
octahedral anions are larger than their tetrahedral counterparts.
Secondly, the pyrazine rings are tilted by differing amounts with
respect to the normal to the 2D layers: those in the octahedral compounds
are significantly more upright. Since the Cu $3d_{x^{2}-y^{2}}$ orbitals
point along the pyrazine directions, these tilting angles might be
expected, to first order, to make little difference to nearest-neighbor
exchange because such rotation is about a symmetry axis as viewed
from the copper site. However, it may be that the different direction
of the delocalized orbitals above and below the rings through which
exchange probably occurs, possibly in conjunction with hybridization
with the anion orbitals, result in an altered next-nearest neighbor
or higher-order interactions, changing the transition temperature.

Within the tetrahedral compounds, the difference in the weighting
of the oscillatory component ($p_{1}+p_{2}$) in \emph{X}$^{-}$~=
BF$_{\text{4}}^{-}$ and ClO$_{\text{4}}^{-}$ probably results from
the difficulty in fitting the fast-relaxing component. Even with little
change in the size of the oscillations, any error assigning the magnitude
of this component will affect the proportion of the $A(t)$ signal
attributed to them. This difficulty is partly due to the resolution-limited
nature of ISIS arising from the pulsed beam structure. In the octahedral
compounds, we found that \emph{X}$^{-}$~= SbF$_{\text{6}}^{-}$
and TaF$_{\text{6}}^{-}$ did not have a resolvable fast-relaxing
component, and consequently $p_{3}$ was set to zero during the fitting
procedure. This is reflected by dashes in the $p_{3}$ and $\lambda_{3}$
columns in \prettyref{tab:Cu(HF2)(pyz)2X-magnetism-parameters}.

The fact that two oscillatory frequencies are observed points to the
existence of at least two magnetically distinct classes of muon site.
In general we find that $p_{1}\approx p_{2}$ for these materials,
making the probability of occupying the sites giving rise to magnetic
precession approximately equal. The weightings $p_{1,2}$ were found
to be significantly less than the weighting $p_{3}$ relating to the
fast-relaxing site. This, in combination with the magnitude of the
fast relaxation $\lambda_{3}(T=0)\gtrsim10\mathrm{\; MHz}$, suggests
that this term should not be identified with the $\frac{1}{3}$-tail
which results from muons with spins parallel to their local field.
(If that were the case then we would expect $(p_{1}+p_{2})/p_{3}=2$,
which we do not observe.) It is likely that each of the components,
$p_{1}$, $p_{2}$ and $p_{3}$, therefore reflect the occurence of
a separate class of muon site in this system. We investigate the possible
positions of these three classes of site in \prettyref{sub:[M(HF2)(pyz)2]X-muon-sites}.

The temperature evolution of the relaxation rates $\lambda_{i}$ is
shown in the right-hand columns of Figs.~\ref{fig:Cu(HF2)(pyz)2X-tetrahedral-magnetism-graphs}
and \ref{fig:Cu(HF2)(pyz)2X-octahedral-magnetism-graphs}. In the
fast-fluctuation limit, the relaxation rates are expected~\cite{hayano1979-muonrelaxation}
to vary as $\lambda\propto\Delta^{2}\tau$, where $\Delta=\sqrt{\gamma_{\murm}^{2}\langle(B-B_{0})^{2}\rangle}$
is the second moment of the local magnetic field distribution (whose
mean is $B_{0}$) in frequency units, and $\tau$ is the correlation
time. In all measured materials, the relaxation rate $\lambda_{1}$,
corresponding to the higher oscillation frequency, starts at a small
value at low temperature and increases as $T_{\mathrm{N}}$ is approached
from below. This is the expected temperature-dependent behavior and
most likely reflects a contribution from critical slowing down of
fluctuations near $T_{\mathrm{N}}$ (described e.g.\ in Ref.~\onlinecite{Pratt2007-critical-behaviour-CoGly}).
In contrast, the relaxation rate $\lambda_{2}$ (associated with the
lower frequency) starts with a higher magnitude at low temperature
and decreases smoothly as the temperature is increased. This is also
the case for the relaxation rate $\lambda_{3}$ of the fast-relaxing
component. This smooth decrease of these relaxation rates with temperature
has been observed previously in magnetic materials~\cite{lancaster2007-YMnO3pressure,lancaster2003-PrO2}
and seems to roughly track the magnitude of the local field. It is
possible that muon sites responsible for $\lambda_{1}$ and $\lambda_{3}$
lie further from the 2D planes than those sites giving rise to $\lambda_{2}$,
and are thus less sensitive to 2D fluctuations, reducing the influence
of any variation in $\tau$. The temperature evolution of $\lambda_{1}$
and $\lambda_{3}$ might then be expected to be dominated by the magnitude
of $\Delta$, which scales with the size of the local field and would
therefore decrease as the magnetic transition is approached from below.
%We note that if $\lambda\propto\Delta^{2}\tau$
%and its variation is dominated by changes in the value of $\Delta$,
%then since $\Delta\propto\nu$ we might expect $\lambda\propto\nu^{2}$.
%However, this relation does not provide a good description of our data. 

The need for nonzero phases $\phi_{i}$ has been identified in previous
studies of molecular magnets~\cite{lancaster2006-CuPzN,Lancaster2004-CuX2(pyz)-and-0DMMs,Lancaster2006-Cu(HCO2)2(pyz),lancaster2007-CuPz2ClO42},
but never satisfactorily explained. One possible explanation for these
might be that the muon experiences delayed state formation. However,
we can rule out the simplest model of this as the phases appear not
to correlate with $\nu_{i}$. Such a correlation would be expected
since a delay of $t_{0}$ before entering the precessing state would
give rise to a component of the relaxation function $a_{i}(t)=\cos\left[2\pi\nu_{i}\left(t+t_{0}\right)\right]=\cos\left(2\pi\nu_{i}t+\phi_{i}\right)$,
with $\phi_{i}\propto\nu_{i}$, which is not observed. This does not
completely rule out delayed state formation, as $t_{0}$ could be
a function of temperature (although this seems unlikely at these temperatures).
Nonzero phases are also sometimes observed when attempting to fit
data with cosinusoidal relaxation functions from systems having incommensurate
magnetic structures. The phase then emerges as an artifact of fitting,
as a cosine with a $\frac{\pi}{4}$ phase shift approximates the zeroth-order
Bessel function of the first kind $J_{0}(\omega t)$ which is obtained
from \muSR\ of an incommensurately-ordered system~\cite{Major1986-ZF-muSR-Cr,Amato1997-muSR-heavy-fermions}.
The Bessel function arises because the distribution of fields seen
by muons at sites is asymmetric. %Time-domain \muSR\  data represents a cosine
%Fourier transform of the field distribution at muon sites within the
%sample, and taking a Fourier transform of this distribution in particular
%gives rise to a Bessel function. 
However, attempts to fit the data with a pair of damped Bessel functions
%\begin{equation}
%A(t)=A_{0}\left(p_{1}J_{0}\left(\omega_{1}t\right)\mathrm{e}^{-\sigma_{1}^{2}t^{2}}+p_{2}J_{0}\left(\omega_{2}t\rig%ht)\mathrm{e}^{-\sigma_{2}^{2}t^{2}}\right)+A_{\mathrm{bg}}\mathrm{e}^{-\lambda_{\mathrm{bg}}t},
%\end{equation}
produced consistently worse fits than fits to \prettyref{eq:M(HF2)(pyz)2X-A(t)-fit},
suggesting that a simple incommensurate structure is not a satisfactory
explanation. It is also possible that several further magnetically-inequivalent
muon sites exist, resulting in multiple, closely-spaced frequencies
which give the spectra a more complex character which is not reflected
in the fitting function. The simpler relaxation function would then
obtain a better fit if the phase were allowed to vary. This has been
observed~\cite{Sugiyama2009-LiCrO2}, for example, in LiCrO$_{\text{2}}$.
A final possibility is that the distribution of fields at muon sites
is asymmetric for another reason, perhaps arising from a complex magnetic
structure. This may give rise to a Fourier transform which is only
able to be fitted with phase-shifted cosines. However, the mechanism
by which this would occur is unclear.

\subsection{Parametrizing exchange anisotropy\label{sub:exchange-anisotropy}}

The extent to which these systems approximate the 2DSLQHA can be quantified
by comparing the transition temperature $T_{\mathrm{N}}$ to the exchange
parameter $J$. The temperature $T_{\mathrm{N}}$ can be extracted
using \muSR , whilst $J$ can be obtained reliably from pulsed-field
magnetization measurements~\cite{goddard2008-2DHMexchange}, heat
capacity or magnetic susceptibility.

%Further, quantum Monte Carlo simulations allow this to be related
%to the interlayer exchange $J_{\perp}$, and the correlation length
%of quantum fluctuations, $\xi$. 

Mean-field theory predicts a simple relationship for the ratio of
the transition temperature $T_{\mathrm{N}}$ and the exchange $J$
given by~\cite{blundell-magnetism-in-condensed-matter-book} 
\begin{equation}
\frac{k_{\mathrm{B}}T_{\mathrm{N}}}{J}=\frac{2}{3}zS(S+1),\label{eq:mean-field-kBTNoverJ}
\end{equation}
 where $k_{\mathrm{B}}$ is Boltzmann's constant, $z$ is the number
of nearest neighbors and $S$ is the spin of the magnetic ions. In
the pseudocubic {[}Cu(HF$_{\text{2}}$)(pyz)$_{\text{2}}${]}\emph{X}
systems, $S=\frac{1}{2}$ and $z=6$, and \prettyref{eq:mean-field-kBTNoverJ}
yields $k_{\mathrm{B}}T_{\mathrm{N}}/J=3$. However, the reduced dimensionality
increases the prevalence of quantum fluctuations, depressing the transition
temperature and in {[}Cu(HF$_{\text{2}}$)(pyz)$_{\text{2}}${]}BF$_{\text{4}}$,
we find $k_{\mathrm{B}}T_{\mathrm{N}}/J\approx0.25$, which is indicative
of large exchange anisotropy.

Combining the experimental measures of $T_{\mathrm{N}}$ and $J$
with the results of quantum Monte Carlo (QMC) simulations allows us
to deduce the exchange anisotropy $J_{\perp}/J$ in the system~\cite{goddard2008-2DHMexchange}.
Specifically, QMC simulations~\cite{yasuda2005-2DHA-JvsTN} for 2DSLQHA
where $10^{-3}\leq J_{\perp}/J\leq1$ are well described by the expression
\begin{equation}
\frac{J_{\perp}}{J}=\mathrm{e}^{b-4\uppi\rho_{\mathrm{s}}/T_{\mathrm{N}}},\label{eq:Yasuda-J-Jprime}
\end{equation}
where $\rho_{\mathrm{s}}$ is the spin stiffness and $b$ is a numerical
constant. For $S=\frac{1}{2}$, the appropriate parameters are $\rho_{\mathrm{s}}/J=0.183$
and $b=2.43$. This expression allows a better estimate of $k_{\mathrm{B}}T_{\mathrm{N}}/J$
in a 3D magnet: evaluating for $J_{\perp}/J=1$ yields $k_{\mathrm{B}}T_{\mathrm{N}}/J=0.95$.
This is lower than the crude mean-field estimate because mean-field
theory takes no account of fluctuations. Estimates of $J$ for our
materials, from pulsed magnetic field studies except where noted,
along with calculated $J_{\perp}/J$ ratios, are shown in the summary
tables throughout this paper.

Another method of parametrizing the exchange anisotropy is to consider
the predicted correlation length of two-dimensional correlations in
the layers at the temperature at which we observe the onset of LRO.
The larger this length, the better isolated the layers can be supposed
to be. This can be estimated by combining an analytic expression for
the correlation length, in a pure 2DSLQHA~\cite{Hasenfratz1991-2DQHA-correlation-length},
$\xi_{\mathrm{2D}}$, with quantum Monte Carlo simulations to obtain
an expression~\cite{Beard1998-2DSLQHA-large-correlation-lengths,Kastner1998-cuprates-review}
appropriate for $1\leq\xi_{\mathrm{2D}}/a\leq350,000$, 
\begin{equation}
\frac{\xi_{\mathrm{2D}}}{a}=0.498\mathrm{e}^{1.131J/k_{\mathrm{B}}T}\left[1-0.44\left(\frac{k_{\mathrm{B}}T}{J}\right)+\mathcal{O}\left(\frac{k_{\mathrm{B}}T}{J}\right)^{2}\right],\label{eq:Beard-correlation-length}
\end{equation}
where $a$ is the square lattice constant, and $T$ is the temperature.
This formula yields $\xi_{\mathrm{2D}}(T_{\mathrm{N}})\approx0.5a$
for the mean-field model ($k_{\mathrm{B}}T_{\mathrm{N}}/J=3$), and
$\xi_{\mathrm{2D}}(T_{\mathrm{N}})\approx a$ for $k_{\mathrm{B}}T_{\mathrm{N}}/J=0.95$
from quantum Monte Carlo simulations (i.e.~\prettyref{eq:Yasuda-J-Jprime}
with $J_{\perp}/J=1$). By comparison, in {[}Cu(HF$_{\text{2}}$)(pyz)$_{\text{2}}${]}BF$_{\text{4}}$
\prettyref{eq:Beard-correlation-length} gives $\xi_{\mathrm{2D}}(T_{\mathrm{N}})\approx50a$,
showing a dramatic increase in the size of correlated regions which
build up in the quasi-2D layers before the onset of LRO.

\subsection{Nonmonotonic field dependence of $T_{\mathrm{N}}$}

\begin{figure*}
\includegraphics{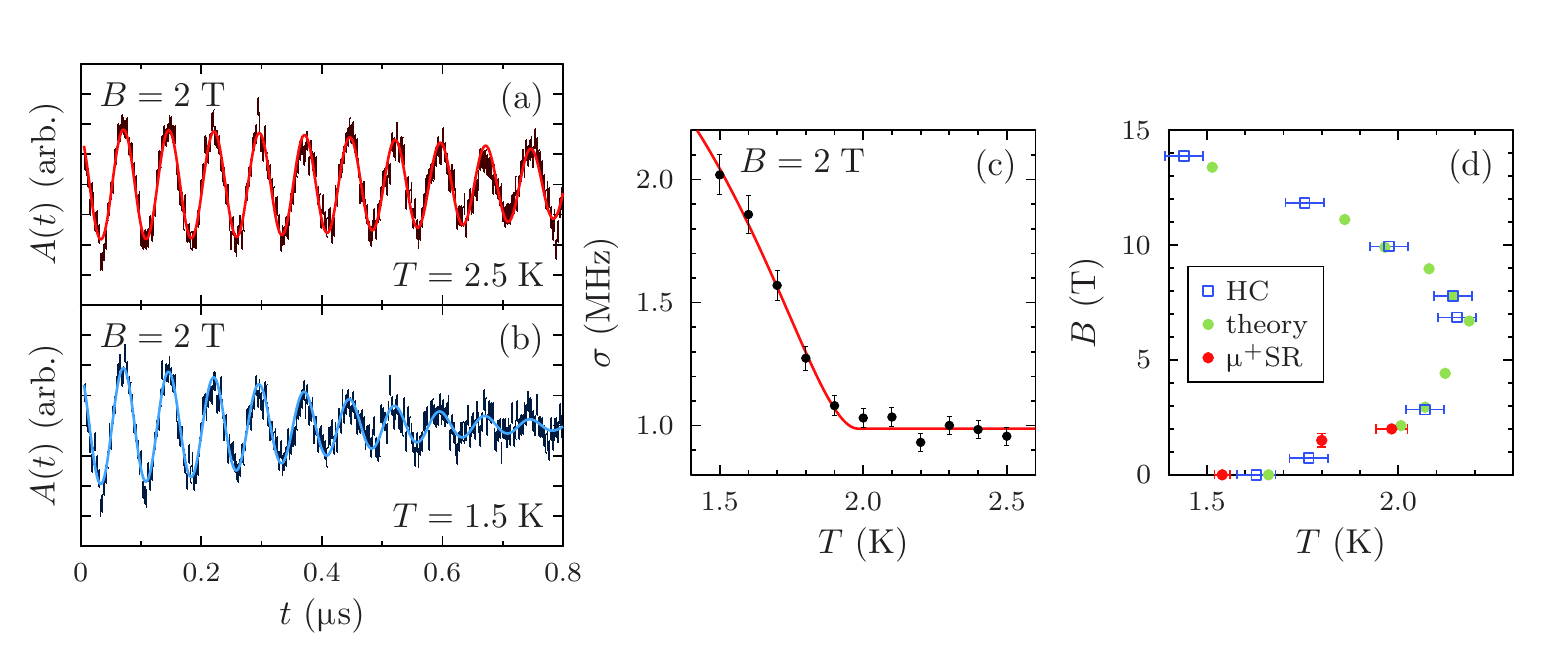}
\caption{Sample TF \muSR\ data measured for {[}Cu(HF$_{\text{2}}$)(pyz)$_{\text{2}}${]}BF$_{\text{4}}$
in an applied field of 2~T are shown in (a) and (b). Data are shown
in the `rotating reference frame', rotating at $\gamma_{\murm}\times1.9\mathrm{\; T}=\mathrm{257\; MHz}$,
nearly cancelling out spin precession induced by the $2\mathrm{\; T}$
applied transverse field. (c) The evolution of the magnetic broadening
$\sigma$ with $T$, showing a magnetic transition at 1.98~K in 2
T. (d) The $B$--$T$ phase diagram from Ref.~\onlinecite{Sengupta2009-nonmonotonic-TN-B-CuHF2pyz2BF4}
showing the nonmonotonic behaviour at low applied magnetic field.
In the key, HC is heat capacity, theory represents the results of
computational modelling, and \muSR\  shows our results from TF measurements
(see main text). \label{fig:nonmonotonic-TN(B)-BF4}}
\end{figure*}

Although we expect the interplane exchange coupling $J_{\perp}$ to
have a large amount of control of the thermodynamic properties of
these materials, it may be the case that single-ion anisotropies are
also responsible for deviations in the behaviour of our materials
from the predictions of the 2DSLQHAF model. In particular, these anisotropies
has been demonstrated to show a crossover to magnetic behaviour consistent
with the 2D $XY$ model~\cite{Xiao2009-XY-q2DHA}. It was recently
reported~\cite{Sengupta2009-nonmonotonic-TN-B-CuHF2pyz2BF4} that
{[}Cu(HF$_{\text{2}}$)(pyz)$_{\text{2}}${]}BF$_{\text{4}}$ exhibits
an unusual nonmonotonic dependence of $T_{\mathrm{N}}$ as a function
of applied magnetic field $B$ {[}see \prettyref{fig:nonmonotonic-TN(B)-BF4}(d){]}.
This behavior was explained as resulting from the small $XY$-like
anisotropy of the spin system in these systems. The physics of the
unusual field-dependence then arises due to the dual effect of $B$
on the spins, both suppressing the amplitude of the order parameter
by polarizing the spins along a given direction, and also reducing
the phase fluctuations by changing the order parameter phase space
from a sphere to a circle. A more detailed explanation for the behavior~\cite{Sengupta2009-nonmonotonic-TN-B-CuHF2pyz2BF4}
reveals that the energy scales of the physics are controlled by a
Kosterlitz--Thouless-like mechanism, along with the interlayer exchange
interaction $J_{\perp}$.

The measurement of the $B$--$T$ phase diagram in {[}Cu(HF$_{\text{2}}$)(pyz)$_{\text{2}}${]}BF$_{\text{4}}$
reported in Ref.~\onlinecite{Sengupta2009-nonmonotonic-TN-B-CuHF2pyz2BF4}
was made by observing a small anomaly in specific heat. In order to
test whether the phase boundary could be determined using muons, we
carried out transverse-field (TF) \muSR\ measurements using the
LTF instrument at S\murm S. In these measurements, the field is applied
perpendicular to the initial muon spin direction, causing a precession
of the muon-spins in the sum of the applied and internal field directed
perpendicular to the muon-spin orientation. Example TF spectra measured
in a field of $2\mathrm{\; T}$ are shown in \prettyref{fig:nonmonotonic-TN(B)-BF4}~(a)
and (b). We find that the spectra are well described by a function
\begin{equation}
A(t)=A(0)\mathrm{e}^{-\sigma^{2}t^{2}/2}\cos(2\pi\nu t+\phi),
\end{equation}
where the phase factor depends on the details of the detector geometry,
and $\sigma$ is proportional to the second moment of the internal
field distribution via $\sigma^{2}=\gamma_{\murm}^{2}\langle B^{2}\rangle$.
Upon cooling through $T_{\mathrm{N}}$ we see a large increase in
$\sigma$, as shown in \prettyref{fig:nonmonotonic-TN(B)-BF4}~(c).
This approximately resembles an order parameter, and we identify the
discontinuity at the onset of the increase with $T_{\mathrm{N}}$
by fitting $\sigma$ with the above-$T_{\mathrm{N}}$ relaxation adding
in quadrature to the additional relaxation present below the transition.
The resulting point at $T_{\mathrm{N}}(B=2\mathrm{\; T})=1.98(4)\mathrm{\; K}$
is shown to be consistent with the predicted low-field phase boundary
in \prettyref{fig:nonmonotonic-TN(B)-BF4}~(d). A further point,
identifiable by its vertical rather than horizontal error bar, was
found by performing a field scan at a fixed temperature of $T=1.8\mathrm{\; K}$.
The field-dependence of the relaxation rate shows a sharp increase
at the transition, at $B=1.5\pm0.3\mathrm{\; T}$.

Points derived from \muSR\  measurements possibly lie slightly lower
in $T$ than both that predicted by theory, and the line predicted
on the basis of the specific heat measurements. The theoretical calculations
use $J/k_{\mathrm{B}}=5.9\mathrm{\; K}$ and $J_{\perp}/J=2.5\times10^{-3}$,
whilst our estimates suggest $J/k_{\mathrm{B}}=6.3\mathrm{\; K}$
and $J_{\perp}/J=0.9\times10^{-3}$. Performing these calculations
for a purely 2D system results in the entire curve shifting to the
left~\cite{Sengupta2009-nonmonotonic-TN-B-CuHF2pyz2BF4}, and consequently
the leftward shift of our data points is consistent with our finding
of increased exchange anisotropy. It is clear that the TF \muSR\ technique
may be used in future to measure the $B$--$T$ phase diagram and
enjoys some of the same advantages it has in ZF over specific heat
and susceptibility in anisotropic systems.

\subsection{Muon response for $T>T_{\mathrm{N}}$\label{sub:[Cu(HF2)(pyz)2]X-paramagnetic}}

\begin{table}
\begin{tabular}{cccccc}
\toprule 
\emph{M} & \emph{X} & $r_{\text{\murm--F}}$ (nm)  & $p_{1}$ (\%)  & $\sigma$ (MHz)  & $T$ (K)\tabularnewline
\midrule 
Cu & BF$_{\text{4}}$  & $0.1038(1)$  & $77(1)$  & $0.29(1)$  & $5.1$\tabularnewline
Cu & ClO$_{\text{4}}$  & $0.1081(2)$  & $70(1)$  & $0.37(1)$  & $5.1$\tabularnewline
Cu & PF$_{\text{6}}$  & $0.1044(2)$  & $74(2)$  & $0.29(2)$  & $5.2$\tabularnewline
Cu & AsF$_{\text{6}}$  & $0.1043(3)$  & $78(2)$  & $0.31(3)$  & $4.9$\tabularnewline
Cu & SbF$_{\text{6}}$  & $0.1047(2)$  & $64(1)$  & $0.30(1)$  & $5.0$\tabularnewline
\midrule 
Cu & BF$_{\text{4}}$  & $0.1042(1)$  & $76(1)$  & $0.26(1)$  & $26$\tabularnewline
Cu & ClO$_{\text{4}}$  & $0.1087(1)$  & $71(1)$  & $0.37(1)$  & $25$\tabularnewline
Cu & SbF$_{\text{6}}$  & $0.1080(3)$  & $59(1)$  & $0.26(1)$  & $30$\tabularnewline
Cu & NbF$_{\text{6}}$  & $0.1039(4)$  & $69(3)$  & $0.32(3)$  & $32$\tabularnewline
Cu & TaF$_{\text{6}}$  & $0.1039(2)$  & $78(2)$  & $0.26(3)$  & $32$\tabularnewline
\midrule
Ni & SbF$_{\text{6}}$  & $0.1068(4)$  & $60(1)$  & $0.38(1)$  & $19$\tabularnewline
Ni & PF$_{\text{6}}$  & $0.1063(5)$  & $66(2)$  & $0.40(1)$  & $8.4$\tabularnewline
\bottomrule
\end{tabular}

\caption{Muon--fluorine dipole--dipole interaction fitted parameters in the
family {[}\emph{M}(HF$_{\text{2}}$)(pyz)$_{\text{2}}${]}\emph{X},
extracted from fitting data to \prettyref{eq:F-mu}. In addition to
separation by metal ion, Cu compounds are grouped by the temperature
at which the measurement was made: those compounds measured over a
range of temperatures appear in both sections of the table.\label{tab:Cu(HF2)(pyz)2X-muF-parameters}}
\end{table}

Above $T_{\mathrm{N}}$, the character of the measured spectra changes
considerably and we observe lower-frequency oscillations characteristic
of the dipole--dipole interaction between muons and fluorine nuclei~\cite{lancaster2007-muF}.
The Cu$^{\text{2+}}$ electronic moments, which dominate the spectra
for $T<T_{\mathrm{N}}$, are disordered in the paramagnetic regime
and fluctuate very rapidly on the muon time scale. They are therefore
motionally narrowed from the spectra, leaving the muon sensitive to
the quasi-static nuclear magnetic moments.

A muon and nucleus interact via the two-spin Hamiltonian 
\begin{equation}
\hat{H}=\sum_{i>j}\frac{\mu_{0}\gamma_{i}\gamma_{j}\hbar}{4\uppi r^{3}}\left[\bm{S}_{i}\cdot\bm{S}_{j}-3\left(\bm{S}_{i}\cdot\hat{\bm{r}}\right)\left(\bm{S}_{j}\cdot\hat{\bm{r}}\right)\right],\label{eq:dipole-coupling-Hamiltonian}
\end{equation}
 where the spins $\bm{S}_{i,j}$ with gyromagnetic ratios $\gamma_{i,j}$
are separated by the vector $\bm{r}$. This gives rise to a precession
of the muon spin, and the muon-spin polarization along a quantization
axis $z$ varies with time as 
\begin{equation}
D_{z}(t)=\frac{1}{N}\left\langle \sum_{m,n}\left|\left\langle m\left|\sigma_{q}\right|n\right\rangle \right|^{2}\mathrm{e}^{\mathrm{i}\omega_{m,n}t}\right\rangle _{q},
\end{equation}
where $N$ is the number of spin states, $\left|m\right\rangle $
and $\left|n\right\rangle $ are eigenstates of the total Hamiltonian
$\hat{H}$, $\sigma_{q}$ is the Pauli spin matrix corresponding to
the direction $q$, and $\left\langle \right\rangle _{q}$ represents
an appropriately-weighted powder average. The vibrational frequency
of the muon--fluorine bond exceeds by orders of magnitude both the
frequencies observable in a \muSR\ experiment, and the frequency
appropriate to the dipolar coupling in \prettyref{eq:dipole-coupling-Hamiltonian};
the bond length probed via these entangled states is thus time-averaged
over thermal fluctuations. Fluorine is an especially strong candidate
for this type of interaction firstly because it is highly electronegative
causing the positive muon to stop close to fluorine ions, and secondly
because its nuclei are 100\% $^{\text{19}}$F, which has $I=\frac{1}{2}$.

\begin{figure}
\includegraphics{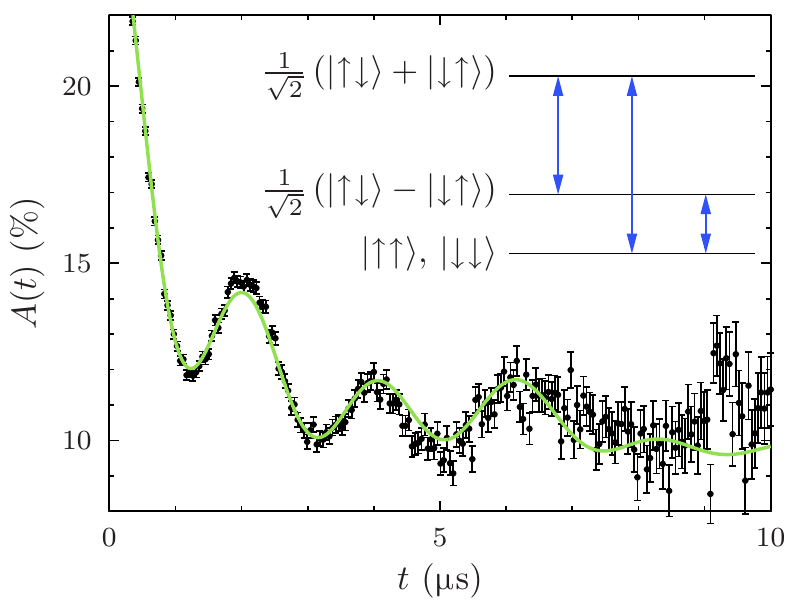} \caption{Data taken at $T=15\mathrm{\; K}\gg T_{\mathrm{N}}=1.4\mathrm{\; K}$
for {[}Cu(HF$_{\text{2}}$)(pyz)$_{\text{2}}${]}BF$_{\text{4}}$,
showing \Fmu\ oscillations along with a fit to \prettyref{eq:F-mu-A(t)}.
The inset shows the energy levels present in a simple system of two
$S=\frac{1}{2}$ spins, along with the allowed transitions.\label{fig:F-mu-BF4}}
\end{figure}

Data were fitted to a relaxation function 
\begin{equation}
A(t)=A_{0}(p_{1}\mathrm{e}^{-\lambda_{\text{F--\murm}}t}D_{z}(t)+p_{2}\mathrm{e}^{-\sigma^{2}t^{2}})+A_{\mathrm{bg}}\mathrm{e}^{-\lambda_{\mathrm{bg}}t},\label{eq:F-mu-A(t)}
\end{equation}
 where the amplitude fraction $p_{1}\approx70\%$ reflects the muons
stopping in a site or set of sites near to a fluorine nucleus, which
result in the observed oscillations $D_{z}(t)$; the weak relaxation
of the muon spins is crudely modelled by a decaying exponential. The
fraction $p_{2}\approx30\%$ describes those muons stopping in a class
of sites primarily influenced by the randomly-orientated fields from
other nuclear moments, giving rise to a Gaussian relaxation with $\sigma\approx0.3\mathrm{\; MHz}$.
Example data and a fit are shown in \prettyref{fig:F-mu-BF4}, whilst
parameters extracted by fitting this function to data from each compound
are shown in \prettyref{tab:Cu(HF2)(pyz)2X-muF-parameters}.

Fits to a variety of different $D_{z}(t)$ functions were attempted,
including that resulting from a simple \Fmu\ bond (previously observed
in some polymers~\cite{lancaster2009-FmuF-PTFE}) and the better-known
\FmuF\ complex comprising a muon and two fluorine nuclei in linear
symmetric configuration, which is seen in many alkali fluorides~\cite{Brewer1986-FmuF}.
This latter model was also modified to include the possibilities of
asymmetric and nonlinear bonds. Previous measurements~\cite{lancaster2007-muF}
made in the paramagnetic regime of {[}Cu(HF$_{\text{2}}$)(pyz)$_{\text{2}}${]}ClO$_{\text{4}}$
suggested that the muon stopped close to a single fluorine in the
HF$_{\text{2}}$ group and also interacted with the more distant proton.
This interaction is dominated by the F--\murm\  coupling and, for
our fitting, the observed muon--fluorine dipole--dipole oscillations
were found to be well described by a single \Fmu\ interaction damped
by a phenomenological relaxation factor. For such \Fmu\ entanglement,
the time evolution of the polarization is described by 
\begin{equation}
D_{z}(t)=\frac{1}{6}\left[1+\sum_{j=1}^{3}u_{j}\cos\left(\omega_{j}t\right)\right],\label{eq:F-mu}
\end{equation}
where $u_{1}=2$, $u_{2}=1$ and $u_{3}=2$. The frequencies $\omega_{j}=j\omega_{\mathrm{d}}/2$,
where $\omega_{\mathrm{d}}=\mu_{0}\gamma_{\murm}\gamma_{\mathrm{F}}\hbar/4\uppi r^{3}$,
in which $\gamma_{\mathrm{F}}=2\pi\times2.518\times10^{8}\mathrm{\; MHz\, T^{-1}}$
is the gyromagnetic ratio of a $^{\text{19}}$F nucleus~\cite{NMR-Encyclopedia2002},
and $r$ is the muon--fluorine separation. These three frequencies
arise from the three transitions between the three energy levels present
in a system of two entangled $S=\frac{1}{2}$ particles (see inset
to \prettyref{fig:F-mu-BF4}). The fact that the relaxation function
is similar in all materials in the series, including {[}Cu(HF$_{\text{2}}$)(pyz)$_{\text{2}}${]}ClO$_{\text{4}}$
which is the only compound studied without fluorine in its anion,
(the only difference being a slight lengthening of the \murm --F
bond, and with no significant change in oscillating fraction) suggests
that the muon site giving rise to the \Fmu\ oscillations in all
systems is near the HF$_{\text{2}}$ bridging ligand.

\begin{figure*}
\includegraphics{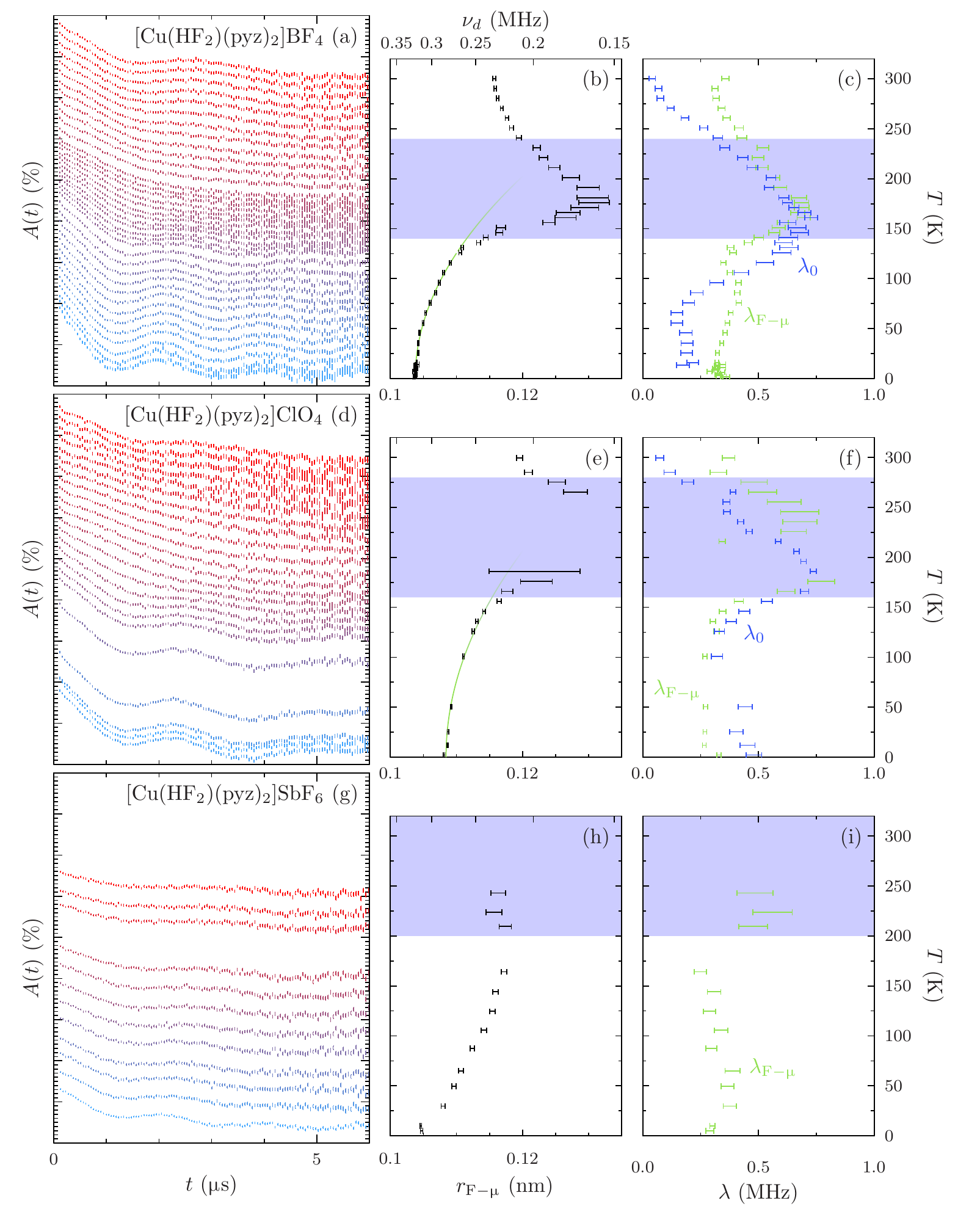} \caption{(a) Muon--fluorine dipole--dipole oscillations in {[}Cu(HF$_{\text{2}}$)(pyz)$_{\text{2}}${]}BF$_{\text{4}}$
over a temperature range $2\leq T\leq300\mathrm{\; K}$. Asymmetry
spectra are displaced vertically so as to approximately align with
the temperature scale on plots (b) and (c). Ticks on the $y$-axis
of (a) denote 1\% asymmetry. Plot (b) shows the fitted value for $r_{\text{F--\murm}}$
as a function of temperature. The line shown is a fit to the low-$T$
points with a $T^{2}$ scaling law, \prettyref{eq:Fmu-r-scaling}.
The upper $x$-axis shows values of the dipole frequency, $\nu_{\mathrm{d}}$,
which correspond to the lower $x$-axis values of $r_{\text{F--\murm}}$.
Plot (c) shows fitted relaxation rates $\lambda_{\text{F--\murm}}$
and $\lambda_{0}$, referring to the relaxation of the \Fmu\ function
$D_{z}(t)$ in \prettyref{eq:F-mu-A(t)}, and the pure relaxation
in \prettyref{eq:F-mu-vanished-A(t)}. Shaded regions indicate temperatures
where $Q=2\uppi\lambda_{\text{F--\murm}}/\omega_{\mathrm{d}}>2$,
roughly parametrizing the disappearance of the oscillations. (d),
(e) and (f) follow (a), (b) and (c), but show data for {[}Cu(HF$_{\text{2}}$)(pyz)$_{\text{2}}${]}ClO$_{\text{4}}$.
(g), (h) and (i) similarly, but for {[}Cu(HF$_{\text{2}}$)(pyz)$_{\text{2}}${]}SbF$_{\text{6}}$
over the range $5\leq T\leq250\mathrm{\; K}$. The $5\mathrm{\; K}$
$A(t)$ plot is omitted because the background is raised substantially
by approach to the transition to LRO.\label{fig:CuHF2pyz2X-Fmu-At-waterfall}}
\end{figure*}

\begin{table}
\begin{tabular}{ccc}
\hline 
material  & $r_{0}\mathrm{\;(nm)}$  & $a\mathrm{\;(10^{-7}\, nm\, K^{-2}})$\tabularnewline
\hline 
{[}Cu(HF$_{\text{2}}$)(pyz)$_{\text{2}}${]}BF$_{\text{4}}$  & $0.10376(3)$  & $3.96(6)$\tabularnewline
{[}Cu(HF$_{\text{2}}$)(pyz)$_{\text{2}}${]}ClO$_{\text{4}}$  & $0.10842$  & $2.7212$\tabularnewline
PVDF~\cite{lancaster2009-FmuF-PTFE}  & $0.10914$  & $1.9488$\tabularnewline
\hline 
\end{tabular}\caption{Fitted values obtained by fitting muon--fluorine bond lengths with
a $T^{2}$ scaling law, as \prettyref{eq:Fmu-r-scaling}.\label{tab:Fmu-r-scaling}}
\end{table}

The temperature evolution of the \Fmu\ signal was studied for $T\leq300\mathrm{\; K}$
in {[}Cu(HF$_{\text{2}}$)(pyz)$_{\text{2}}${]}BF$_{\text{4}}$ and
{[}Cu(HF$_{\text{2}}$)(pyz)$_{\text{2}}${]}ClO$_{\text{4}}$. In
both cases, the dipole--dipole oscillations disappear gradually in
a temperature range $150\lesssim T\lesssim250\mathrm{\; K}$, with
oscillations totally absent in the center of this range, followed
by reappearing as temperature is increased further. Plots of $A(t)$
spectra at a variety of temperatures are shown in \prettyref{fig:CuHF2pyz2X-Fmu-At-waterfall}~(a)
and (d). The data were initially fitted to \prettyref{eq:F-mu-A(t)},
with all parameters left free to vary. The temperature-evolution of
the muon--fluorine bond length, $r_{\text{F--\murm}}$, can be seen
in \prettyref{fig:CuHF2pyz2X-Fmu-At-waterfall}~(b) and (e). The
spectra were also fitted with
\begin{equation}
A(t)=A_{0}(p_{1}\mathrm{e}^{-\lambda_{0}t}+p_{2}\mathrm{e}^{-\sigma^{2}t^{2}})+A_{\mathrm{bg}}\mathrm{e}^{-\lambda_{\mathrm{bg}}t},\label{eq:F-mu-vanished-A(t)}
\end{equation}
a sum of an exponential and a Gaussian relaxation, which might be
expected to describe the data in the region where the oscillations
vanish. Both this relaxation and that extracted from \prettyref{eq:F-mu-A(t)}
are plotted in \prettyref{fig:CuHF2pyz2X-Fmu-At-waterfall}~(c) and
(f), labelled $\lambda_{0}$ and $\lambda_{\text{F--\murm}}$ respectively.

This bond length appears to grow and then shrink by nearly 20\% over
the $100\mathrm{\; K}$ range where the oscillations fade from the
spectra and reappear. This variation is significantly larger than
any variation in crystal lattice parameters which would be expected.
Since the oscillations visibly disappear from the measured spectra,
results from fitting with an oscillatory relaxation function are artifacts
of the fitting procedure: since the frequencies scale with $1/r^{3}$,
increasing bond length together with the associated relaxation rate
fits the data with a suppressed oscillatory signal. This can be approximately
quantified by examining the ratio $Q=2\uppi\lambda_{\text{F--\murm}}/\omega_{\mathrm{d}}$,
where a large value indicates that the function relaxes significantly
before a single \Fmu\ oscillation is completed. The shaded regions
in \prettyref{fig:CuHF2pyz2X-Fmu-At-waterfall} show where $Q>2$,
which acts as an approximate bound on where the parametrization in
\prettyref{eq:F-mu-A(t)} would be expected to fail. In the low-$T$
region where $Q<2$, the bond lengths appear to scale roughly as $T^{2}$,
which has previously been observed in fluoropolymers~\cite{lancaster2009-FmuF-PTFE}.
Parameters extracted from fitting to
\begin{equation}
r_{\text{F--\murm}}=aT^{2}+r_{0}\label{eq:Fmu-r-scaling}
\end{equation}
 are shown in \prettyref{tab:Fmu-r-scaling}.

The observation in these two samples of \Fmu\ oscillations which
disappear and then reappear is puzzling. While we have not identified
a definitive mechanism, we can probably rule out an electronically
mediated effect since, for $T\gg T_{\mathrm{N}}$, the Cu moment fluctuations
will be outside the muon time-window. An explanation could involve
nearby nuclear moments, possibly influenced by a thermally-driven
structural distortion or instability.

A similar study of {[}Cu(HF$_{\text{2}}$)(pyz)$_{\text{2}}${]}SbF$_{\text{6}}$
is shown in \prettyref{fig:CuHF2pyz2X-Fmu-At-waterfall}~(g), (h)
and (i). In this material, the oscillations appear not to vanish over
the temperature range studied, though we cannot rule out a brief disappearance
at $T\approx200\mathrm{\; K}$. Instead, the oscillations show an
apparently monotonic increase in damping with temperature, and the
fitted bond length does not follow \prettyref{eq:Fmu-r-scaling}.
The shaded region in \prettyref{fig:CuHF2pyz2X-Fmu-At-waterfall}~(h)
and (i) has no upper bound, though we cannot rule out a constraint
at $T>250\mathrm{\; K}$. The pure relaxation $\lambda_{0}$ is omitted
because there is no region where the \Fmu\ oscillations are sufficiently
damped for \prettyref{eq:F-mu-vanished-A(t)} to be a good parametrization.

\subsection{Muon site determination\label{sub:[M(HF2)(pyz)2]X-muon-sites}}

\begin{figure}
\includegraphics{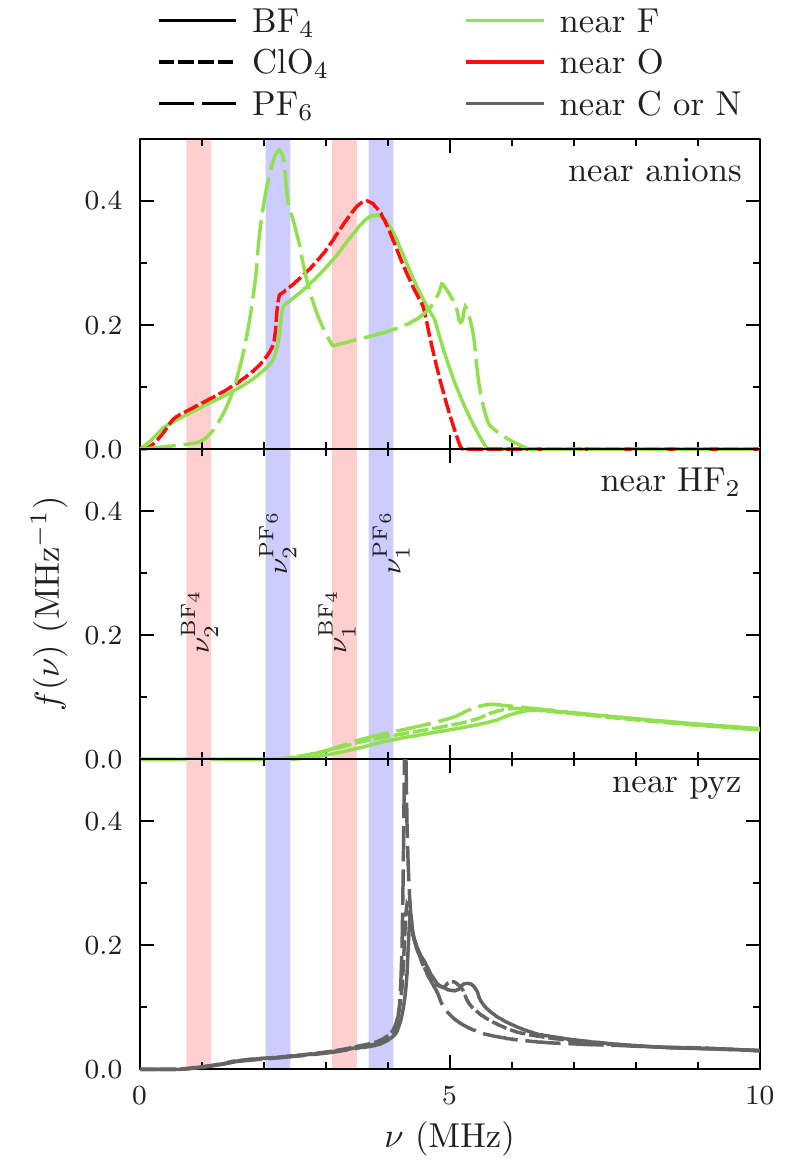}

\caption{Probability density functions of muon precession frequencies at positions
close to likely muon stopping sites in {[}Cu(HF$_{\text{2}}$)(pyz)$_{\text{2}}${]}\emph{X}
with Cu$^{\text{2+}}$ moments $\mu=\mu_{\mathrm{B}}$. The graphs
show dipole fields near the fluorine or oxygen atoms in the negative
anions; near the fluorine atoms in the bifluoride ligands; and near
the carbon and nitrogen atoms, as a proxy for proximity to the pyrazine
ring. The type of line indicates the compound for which the calculation
was performed. The muon site is constrained to be close to particular
atoms, indicated by line color. The shaded areas indicate ranges of
fitted frequencies as $T\rightarrow0$; two frequencies $\nu_{1,2}^{\mathrm{\mathrm{BF_{4}}}}$
represent those observed where \emph{X}$^{-}$~= BF$_{\text{4}}^{-}$,
and $\nu_{1,2}^{\mathrm{PF_{6}}}$ those observed in the \emph{X}$^{-}$~=
PF$_{\text{6}}^{-}$ analogue. \label{fig:2DMM-pdfs}}
\end{figure}

\begin{figure}
\includegraphics{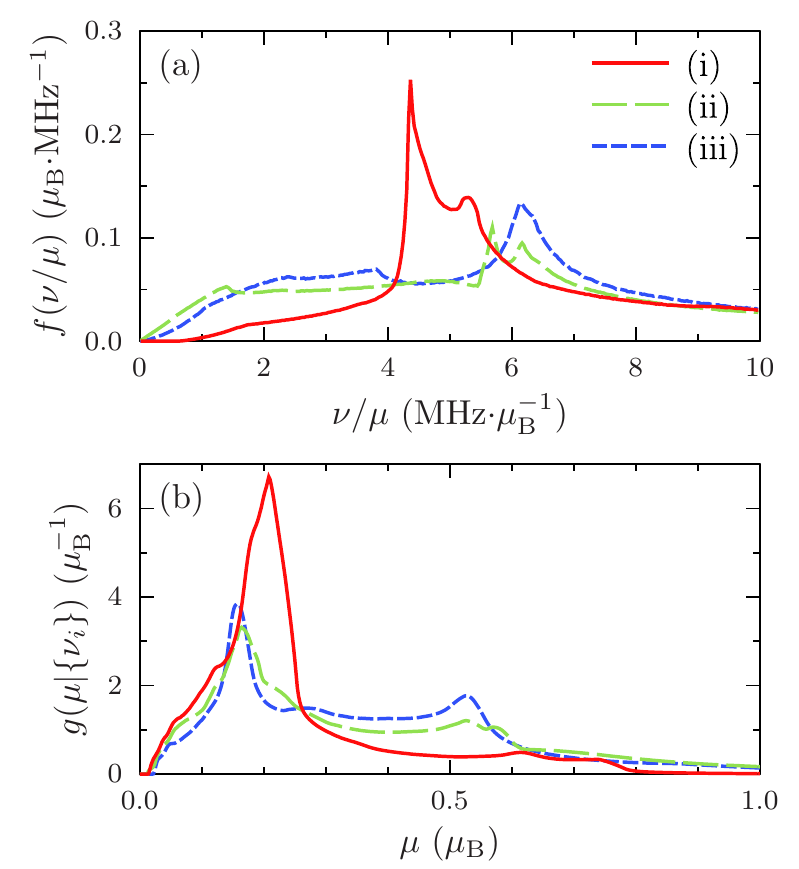}
\caption{Probability density functions for muons in the putative oscillating
sites near the pyrazine rings in {[}Cu(HF$_{\text{2}}$)(pyz)$_{\text{2}}${]}BF$_{\text{4}}$
(a) for muon precession frequency $\nu$ assuming that the moment
on the copper site $\mu=\mu_{\mathrm{B}}$, created fom a histogram
of dipole fields evaluated using \prettyref{eq:dipole-field} at points
satisfying the constraints detailed in the text; and (b) for moment
on the copper sites given the frequencies actually observed, evaluated
using the pdfs in (a) and \prettyref{eq:g(mu)}. Lines represent trial
magnetic structures. All exhibit antiferromagnetic coupling through
both the bifluoride and pyrazine exchange paths, whilst (i) has copper
moments pointing at $45^{\circ}$ to the pyrazine grid, (ii) has moments
along one of the pyrazine grid directions ($\bm{a}$ or $\bm{b}$),
and (iii) has copper moments pointing along the bifluoride axis ($\bm{c}$).\label{fig:mu_Cu-pdf} }
\end{figure}

Combining the data measured above and below the transition in these
materials allows us to attempt to construct a self-consistent picture
of possible muon sites. The observed dipole--dipole observations above
$T_{\mathrm{N}}$ suggest that at least one muon stopping site is
near a fluorine ion. We consider three classes of probable muon site:
Class~I sites near the fluorine ions in the HF$_{\text{2}}$ groups,
Class~II sites near the pyrazine rings, and Class~III sites near
the anions at the centre of the pseudocubic pores. Comparison of Tables~\ref{tab:Cu(HF2)(pyz)2X-magnetism-parameters}
and \ref{tab:Cu(HF2)(pyz)2X-muF-parameters} show that the dominant
amplitude component for $T>T_{\mathrm{N}}$ arises from dipole--dipole
oscillatory component $p_{1}\mathrm{e}^{-\lambda t}D_{z}(t)$ and
from the fast-relaxing component $p_{3}\mathrm{e}^{-\lambda_{3}t}$
for $T<T_{\mathrm{N}}$, and that these are comparable in amplitude.
It is plausible therefore, to suggest that these two signals correspond
to contributions from the same Class~I muon sites near the HF$_{\text{2}}$
groups. Moreover, the analysis of the $T>T_{\mathrm{N}}$ spectra
in the previous section implies that this site lies $r_{\text{\murm--F}}\approx0.11\;\text{nm}$
from an F in the HF$_{\text{2}}$ groups. The remainder of the signal
(the oscillating fraction below $T_{\mathrm{N}}$ and the Gaussian
relaxation above) can also be identified, suggesting that the sites
uncoupled from fluorine nuclei (Classes~II and/or III) result in
the magnetic oscillations observed for $T<T_{\mathrm{N}}$.

We note further that the evidence from \Fmu\ oscillations makes
the occurence of Class~III muon sites unlikely. The fact that spectra
observed for $T>T_{\mathrm{N}}$ in the \emph{X}$^{-}$~= ClO$_{\text{4}}^{-}$
material are nearly identical to those in all other compounds, in
which \emph{X} contains fluorine, suggest that the muons do not stop
near the anions. Moreover, as discussed in \prettyref{sec:[Cu(pyz)2(pyo)2]X2}
and \ref{sec:[Cu(pyo)6]X2} below, we observe no \Fmu\ oscillations
in {[}Cu(pyz)$_{\text{2}}$(pyo)$_{\text{2}}${]}\emph{X}$_{\text{2}}$
(\emph{Y}$^{-}$~= BF$_{\text{4}}^{-}$, PF$_{\text{6}}^{-}$) or
{[}Cu(pyo)$_{\text{6}}${]}(BF$_{\text{4}}$)$_{\text{2}}$, suggesting
that muons do not stop preferentially near these fluorine-rich anions
either. We therefore rule out the existence of Class~III muon sites
and propose that the magnetic oscillations measured for $T<T_{\mathrm{N}}$
most probably arise due to Class~II sites found near the pyrazine
ligands.

Below $T_{\mathrm{N}}$, the measured muon precession frequencies
allow us to determine the magnetic field at these Class~II muon sites
via $\nu=\gamma_{\murm}B/2\pi$. Simulating the magnetic field inside
the crystal therefore allows us to compare these $B$-fields with
those predicted for likely magnetic structures and may permit us to
constrain the ordered moment. For the case of our ZF measurements
in the antiferromagnetic state, the local magnetic field at the muon
site $\bm{B}_{\mathrm{local}}$ is given by 
\begin{equation}
\bm{B}_{\mathrm{local}}=\bm{B}_{\mathrm{dipole}}+\bm{B}_{\mathrm{hyperfine}},
\end{equation}
 where $\bm{B}_{\mathrm{dipole}}$ is the dipolar field from magnetic
ions located within a large sphere centred on the muon site and $\bm{B}_{\mathrm{hyperfine}}$
the contact hyperfine field caused by any spin density overlapping
with the muon wavefunction. This spin density is difficult to estimate
accurately, particularly in complex molecular systems, but it is probable
for insulating materials such as these that the spin density on the
copper ion is well localised and so we ignore the hyperfine contribution
in our analysis. %if there are moments on the
%pyrazine rings then the delocalisation of the $\uppi$-orbitals may
%result in spin-density overlapping with the muon site.
The dipole field $\bm{B}_{\mathrm{dipole}}$ is a function of the
coordinate of the muon site $\bm{r}_{\murm}$, and comprises a vector
sum of the fields from each of the magnetic ions in the crystal approximated
as a point dipole, so that 
\begin{equation}
\bm{B}_{\mathrm{dipole}}(\bm{r}_{\murm})=\mu\sum_{i}\frac{\mu_{0}}{4\uppi r^{3}}\left[3(\hat{\bm{\mu}}_{i}\cdot\hat{\bm{r}})\hat{\bm{r}}-\hat{\bm{\mu}}_{i}\right],\label{eq:dipole-field}
\end{equation}
 where $\bm{r}=\bm{r}_{i}-\bm{r}_{\murm}$ is the relative position
of the muon and the $i^{\text{th}}$ ion with magnetic moment $\bm{\mu}_{i}=\mu\hat{\bm{\mu}}_{i}$,
and $i$ is an index implying summation over all of the ions which
make up the crystal. %This sum is made tractable
%because the $1/r^{3}$-dependence of a dipole field falls off faster
%than the $r^{2}$ increase in surface area of a sphere, meaning that
%the value converges and an accurate estimate can be performed with
%a sufficiently large Lorentz sphere of dipoles~\cite{Blundell2009-dipolefielddistributions}.

Although these materials are known to be antiferromagnetic from their
negative Curie--Weiss temperatures and zero spontaneous magnetization
at low temperatures~\cite{Manson2009-Cu-SbF6,manson2006-CuHF2pyz2BF4-chemcomm},
their magnetic structures are unknown. Dipole field simulations were
therefore performed for a variety of trial magnetic structures with
$\mu=\mu_{\mathrm{B}}$. %Two effects determine the magnitude of the dipole field: for sites
%near to a magnetic ion, the primary effect is the
%distance from that ion; further from a dipole, its
%rapidly dropping contribution becomes comparable to those of neighboring
%ions which can lead to partial or complete cancellation
%of the dipolar field~\cite{Blundell2010-EuO}.
%The magnetic structure  the direction of the
%copper moments; (b) the coupling via a pyrazine ring (assumed to be
%the same along both axes due to the symmetry of the crystal structure);
%and (c) the coupling via the HF$_{\text{2}}$ ligand.
We analyse the results of these calculations using a probabilistic
method. We begin by allowing the possibility that the magnetic precession
signal could arise from any of the possible classes of muon site identified
above. Random positions in the unit cell were generated and dipole
fields calculated at these. To prevent candidate sites lying too close
to atoms we constrain all sites such that $r_{\murm\text{--\emph{A}}}>0.1\mathrm{\; nm}$
where $A$ is any atom. Possible Class~I muon sites were identified
with $r_{\text{\murm--F}}=r_{0}\pm0.01\mathrm{\; nm}$ (where $r_{0}$
is the muon--fluorine distance established from \Fmu\ oscillations)
and possible Class~II sites were selected with the constraint that
$0.10\leq r_{\text{\murm--C,N}}\leq0.12\mathrm{\; nm}$. %Similarly, in oxides,
%muons have been shown to stop around $0.1\;\mathrm{nm}$ from an O$^{\text{2}-}$
%ion~\cite{Brewer1991-muon-1A-from-O}, isolating sites close to the
%ClO$_{\text{4}}$ anion where . 
%Since dipole fields are linearly
%proportional to the moment, it is then permissible to scale the fields
%observed to explore the effect of changing this value. 
The predicted probability density function (pdf) of muon precession
frequencies (resulting from the magnitudes of the calculated fields)
are plotted in \prettyref{fig:2DMM-pdfs}, with the observed frequencies
superimposed. Results are shown for a trial magnetic structure comprising
copper spins lying in the plane of the pyrazine layers and at $45^{\circ}$
to the directions of the pyrazine chains, and with spins arranged
antiferromagnetically both along those chains and along the HF$_{\text{2}}$
groups. This candidate structure is motivated by analogy with {[}Cu(pyz)$_{\text{2}}${]}(ClO$_{\text{4}}$)$_{\text{2}}$,
which also comprises Cu$^{\text{2+}}$ ions in layers of 2D pyrazine
lattices~\cite{Tsyrulin2010-Cu(pz)2(ClO4)2}, and with the parent
phases of the cuprate superconductors~\cite{Kastner1998-cuprates-review},
which are also two-dimensional Heisenberg systems of $S=\frac{1}{2}$
Cu$^{\text{2+}}$ ions. Other magnetic structures investigated give
qualitatively similar results. From \prettyref{fig:2DMM-pdfs} it
is clear that the only sites with significant probability density
near to the observed frequencies are those lying near the anions (i.e.\ Class~III
sites) which we have argued are not compatible with our data. The
more plausible muon sites correspond to higher frequencies than those
observed. Our conclusion is that it is likely that the Cu$^{\text{2+}}$
moments are rather smaller than the $\mu_{\mathrm{B}}$ assumed in
the initial calculation.

\begin{figure}
\includegraphics{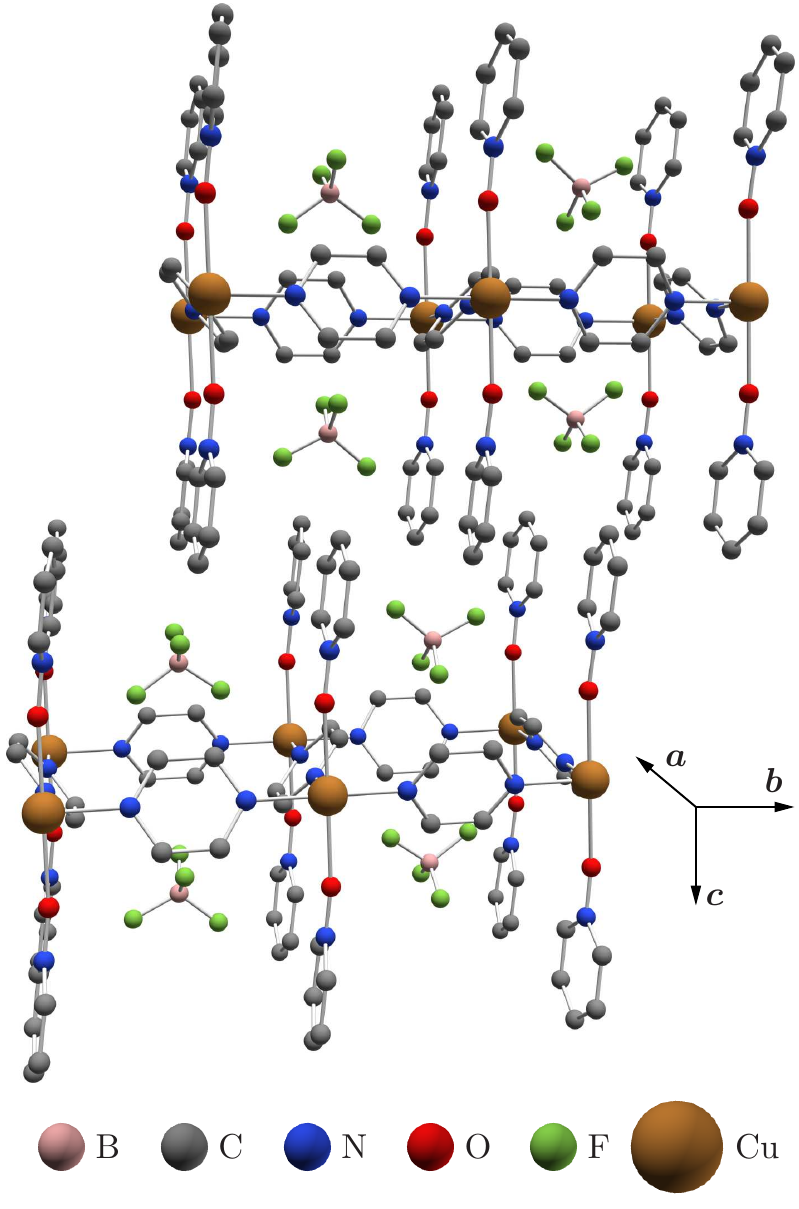}
\caption{Structure of {[}Cu(pyz)$_{\text{2}}$(pyo)$_{\text{2}}${]}(BF$_{\text{4}}$)$_{\text{2}}$.
Copper ions lie in 2D square layers, bound by pyrazine rings. Pyridine-\emph{N}-oxide
ligands protrude from the coppers in a direction approximately perpendicular
to these layers. Tetrafluoroboride ions fill the pores remaining in
the structure. Ion sizes are schematic; copper ions are shown twice
as large for emphasis, and hydrogens have been omitted for clarity.\label{fig:Cu(pyz)2(pyo)2(BF4)2-structure}}
\end{figure}

If we accept that the muon sites giving rise to magnetic precession
are near the pyrazine groups then we may use this calculation to constrain
the size of the copper moment. Since $\nu$ is obtained from experiment,
what we would like to know is $g(\mu\vert\nu)$, the pdf of copper
moment $\mu$ \textit{given} the observed $\nu$. This can be obtained
from our calculated $f(\nu/\mu)$ using Bayes' theorem \cite{Sivia2006-Bayes},
which yields 
\begin{equation}
g(\mu\vert\nu)=\frac{\frac{1}{\mu}f(\nu/\mu)}{\int_{0}^{\mu_{{\rm max}}}\frac{1}{\mu'}f(\nu/\mu')\,{\rm d}\mu'}\,,
\end{equation}
 where we have assumed a prior probability for the copper moment that
is uniform between zero and $\mu_{{\rm max}}$. We take $\mu_{{\rm max}}=2\mu_{{\rm B}}$,
although our results are insensitive to the precise value of $\mu_{{\rm max}}$
as long as it is reasonably large. When multiple frequencies $\{\nu_{i}\}$
are present in the spectra, it is necessary to multiply their probabilities
of observation in order to obtain the chance of their simultaneous
observation, so we evaluate 
\begin{equation}
g(\mu\vert\{\nu_{i}\})\propto\prod_{i}\int_{\nu_{i}-\Delta\nu_{i}}^{\nu_{i}+\Delta\nu_{i}}f(\nu_{i}/\mu)\,{\rm d}\nu_{i}\,,\label{eq:g(mu)}
\end{equation}
 where $\Delta\nu_{i}$ is the error on the fitted frequency. Results
are shown in \prettyref{fig:mu_Cu-pdf}, along with the dipole field
pdfs which gave rise to them. By inspection of the pdfs, the copper
moment is likely to be $\mu\lesssim0.5\mu_{\mathrm{B}}$. The dipole
field simulations results also lend weight to our contention that
the oscillatory signal cannot arise from the sites that also lead
to the \Fmu\ component above-$T_{\mathrm{N}}$. If this were the
case then the most likely moment on the copper would be $\mu_{\mathrm{Cu}}\lesssim0.2\mu_{\mathrm{B}}$,
which seems unreasonably small. We note that moment sizes of $\mu\lesssim0.5\mu_{\mathrm{B}}$
were also observed for the 2DSLQHA system La$_{\text{2}}$CuO$_{\text{4}}$
(a recent estimate~\cite{Reehuis2006-neutron-high-B-La2CuO4} from
neutron diffraction gave $[0.42\pm0.01]\mu_{\mathrm{B}}$), despite
the predictions of $0.6\mu_{\mathrm{B}}$ from spin wave theory and
Quantum Monte Carlo~\cite{Chakravarty1989-2DHA-moment-size}. It
was suggested in that case~\cite{Chakravarty1989-2DHA-moment-size}
that disorder might play a role in reducing the moment sizes; an additional
possible mechanism for this suppression is ring exchange~\cite{Motrunich2005-ring-exchange-triangular-lattice,Katanin2002-ring-exchange-La2CuO4}.

\begin{figure*}
\includegraphics{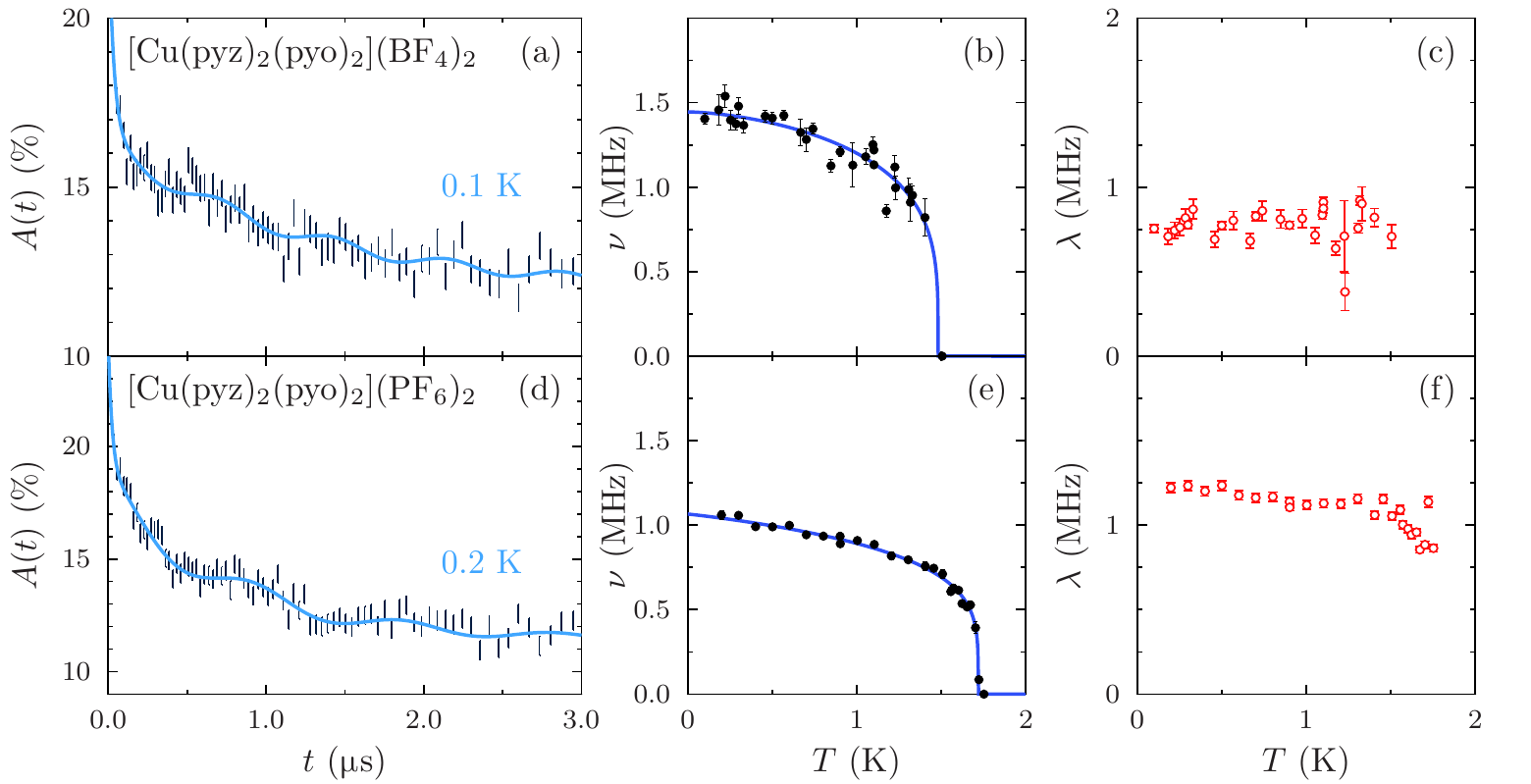}
\caption{Example data and fits for {[}Cu(pyz)$_{\text{2}}$(pyo)$_{\text{2}}${]}\emph{Y}$_{\text{2}}$.
From left to right: (a) and (d) show sample asymmetry spectra $A(t)$
for $T<T_{\mathrm{N}}$ along with a fit to \prettyref{eq:Cu(pyz)2(pyo)2X2-A(t)};
(b) and (e) show the frequency $\nu$ as a function of temperature;
and (c) and (f) show relaxation rates $\lambda_{i}$ as a function
of temperature. The relaxation rates $\lambda_{1}$, associated with
the oscillation, and $\lambda_{3}$, the fast-relaxing initial component,
do not vary significantly with $T$ and are not shown. Only a slight
trend in $\lambda_{2}$ in the \emph{Y}$^{-}$~=~PF$_{\text{6}}^{-}$
material {[}graph (f){]} is observed.\label{fig:Cu(pyz)2(pyo)2X2-magnetism-graphs}}
\end{figure*}

One limitation of this analysis is that the mechanism for magnetic
coupling of copper ions through the pyrazine rings is postulated to
be via spin exchange, in which small magnetic polarisations are induced
on intervening atoms~\cite{Lloret1998-spin-pol-pyz-pyd}. Density
functional theory calculations estimate that these are small, with
the nitrogen and carbon moments estimated at $\mu_{\mathrm{C}}\approx0.01\mu_{\mathrm{B}}$
and $\mu_{\mathrm{N}}\approx0.07\mu_{\mathrm{B}}$, respectively~\cite{Middlemiss2008-CuHF2pyz2BF4-DFT}.
However, their effect may be non-negligible: they may be significantly
closer to the muon site than a copper moment, and dipole fields fall
off rapidly, as $1/r^{3}$. Further, since much of the electron density
in a pyrazine ring is delocalised in $\uppi$-orbitals, the moments
may not be point-like, as assumed in our dipole field calculations.
Further, this may lead to overlap of spin density at the muon site
and result in a nonzero hyperfine field. %Further, muons are effectively a defect in the crystal
%lattice and distort their local environment accordingly; using the
%unperturbed crystal structure is therefore an approximation. 

\section{$\text{[Cu(pyz)}_{\text{2}}\text{(pyo)}_{\text{2}}\text{]\emph{Y}}_{\text{2}}$\label{sec:[Cu(pyz)2(pyo)2]X2}}

In this section, we report the magnetic behavior of another family
of molecular systems which shows quasi-2D magnetism, but for which
the interlayer groups are very different and arranged in a completely
different structure, resulting in a 2D coordination polymer. This
system is {[}Cu(pyz)$_{\text{2}}$(pyo)$_{\text{2}}${]}\emph{Y}$_{\text{2}}$,
where \emph{Y}$^{-}$~= BF$_{\text{4}}^{-}$, PF$_{\text{6}}^{-}$.
As with the previous case, $S=\frac{1}{2}$ Cu$^{\text{2+}}$ ions
are bound in a 2D square lattice of {[}Cu(pyz)$_{\text{2}}${]}$^{\text{2+}}$
sheets lying in the $ab$-plane. Pyridine-\emph{N}-oxide (pyo) ligands
{[}shown in \prettyref{fig:ligands}~(b){]} protrude from the copper
ions along the $c$-direction, perpendicular to the $ab$-plane in
the \emph{Y}$^{-}$~= PF$_{\text{6}}^{-}$ material, but making an
angle $\beta-90\approx29^{\circ}$ with the normal in \emph{Y}$^{-}$~=
BF$_{\text{4}}^{-}$. The anions then fill the pores remaining in
the structure. The structure of {[}Cu(pyz)$_{\text{2}}$(pyo)$_{\text{2}}${]}(BF$_{\text{4}}$)$_{\text{2}}$
is shown in \prettyref{fig:Cu(pyz)2(pyo)2(BF4)2-structure}.

In a typical synthesis, an aqueous solution of Cu\emph{Y}$_{\text{2}}$
hydrate (\emph{Y$^{-}$}~= BF$_{\text{4}}^{-}$ or PF$_{\text{6}}^{-}$)
was combined with an ethanol solution that contained a mixture of
pyrazine and pyridine-\emph{N}-oxide or 4-phenylpyridine-\emph{N}-oxide.
Deep blue-green solutions were obtained in each case, and when allowed
to slowly evaporate at room temperature for a few weeks, dark green
plates were recovered in high yield. Crystal quality could be improved
by sequential dilution and collection of multiple batches of crystals
from the original mother liquor. The relative amounts of pyz and pyo
were optimized in order to prevent formation of compounds such as
Cu\emph{Y}$_{\text{2}}$(pyz)$_{\text{2}}$ or {[}Cu(pyo)$_{\text{6}}${]}\emph{Y}$_{\text{2}}$.

Samples were measured in the LTF apparatus at S\murm S. Example data
measured on {[}Cu(pyz)$_{\text{2}}$(pyo)$_{\text{2}}${]}\emph{Y}$_{\text{2}}$
are shown in Fig.~\ref{fig:Cu(pyz)2(pyo)2X2-magnetism-graphs}, where
we observe oscillations in $A(t)$ at a single frequency below $T_{\mathrm{N}}$.
Data were fitted to a relaxation function 
\begin{equation}
A(t)=A_{0}\left(p_{1}\cos(2\uppi\nu_{1}t)\mathrm{e}^{-\lambda_{1}t}+p_{2}\mathrm{e}^{-\lambda_{2}t}+p_{3}\mathrm{e}^{-\lambda_{3}t}\right)+A_{\mathrm{bg}}.\label{eq:Cu(pyz)2(pyo)2X2-A(t)}
\end{equation}
The small amplitude fraction $p_{1}<10\%$ for both samples refers
to muons stopping in a site or set of sites with a narrow distribution
of quasi-static local magnetic fields, giving rise to the oscillations;
$p_{2}\approx50\%$ is the fraction of muons stopping in a class of
sites giving rise to a large relaxation rate $30\lesssim\lambda\lesssim60$~MHz
and $p_{3}\approx50\%$ represents the fraction of muons stopping
in sites with a small relaxation rate $\lambda_{3}\approx1\;\mathrm{MHz}$.
The data from these compounds fit best with $\phi=0$, and it is thus
omitted from this expression. Frequencies obtained from fitting the
data to \prettyref{eq:Cu(pyz)2(pyo)2X2-A(t)} were then modelled with
\prettyref{eq:nuT1-T-over-TN-alpha-beta}. The results of these fits
are shown \prettyref{fig:Cu(pyz)2(pyo)2X2-magnetism-graphs}, and
\prettyref{tab:other2dmm-params}.

Our results show that {[}Cu(pyz)$_{\text{2}}$(pyo)$_{\text{2}}${]}(BF$_{\text{4}}$)$_{\text{2}}$
has a transition temperature $T_{\mathrm{N}}=1.5\pm0.1\mathrm{\; K}$
and a quasi-static magnetic field at the muon site $\nu_{1}(T=0)=1.4\pm0.1\mathrm{\; MHz}$.
No quantities other than $\nu_{1}$ show a significant trend in the
temperature region $0.1\leq T\leq1.6\mathrm{\: K}$. Above the transition,
purely relaxing spectra are observed, displaying no \Fmu\ oscillations.
As suggested above, this makes the existence of muon sites near the
anions unlikely. %Measurements made using the EMU spectrometer in the temperature range
%$1.4\leq T\leq4.1\mathrm{\: K}$ do not show any significant variation
%for a variety of attempted fitting functions, suggesting that the
%spectra obtained are indistinguishable and therefore that there are
%no trends detectable to muons in this region. 
We find a critical exponent of $\beta=0.25\pm0.10$, where the large
uncertainty results in part from the difficulty in fitting the $A(t)$
data in the critical region.

Our results for {[}Cu(pyz)$_{\text{2}}$(pyo)$_{\text{2}}${]}(PF$_{\text{6}}$)$_{\text{2}}$
show that the transition temperature is slightly higher at $T_{\mathrm{N}}=1.72\pm0.02\mathrm{\; K}$
and the oscillations occur at a lower frequency of $\nu_{1}(T=0)=1.07\pm0.03\mathrm{\; MHz}$.
The relaxation rates $\lambda_{2}$ and $\lambda_{3}$ also decrease
in magnitude as temperature is increased, settling on roughly constant
values $\lambda_{2}\approx0.6\mathrm{\; MHz}$ and $\lambda_{3}\approx15\mathrm{\; MHz}$
for $T>T_{\mathrm{N}}$. No other quantities show a significant trend
in the temperature region $0.2\leq T\leq1.7\mathrm{\: K}$. Above
the transition, relaxing spectra devoid of \Fmu\ oscillations are
again observed. The critical exponent $\beta=0.22\pm0.02$.

The small amplitude of the oscillations, common to both samples, might
be explained in a number of ways. The materials may undergo long-range
ordering but there may be an increased likelihood of stopping in sites
where the magnetic field nearly precisely cancels. Alternatively,
a range of similar muon sites may be present with a large distribution
of frequencies, or alternatively the presence of dynamics, washing
out any clear oscillations in large fractions of the spectra and instead
resulting in a relaxation. Finally, we cannot exclude the possibility
that only a small volume of the sample undergoes a magnetic transition;
this may indicate the presence of a small impurity phase, possibly
located at either grain boundaries or, given that this is a powder
sample, near the crystallites' surfaces. We note also that the behavior
of fitted parameters in these materials is qualitatively similar to
that reported in CuCl$_{\text{2}}$(pyz), where there is also a relatively
small precessing fraction of muons and little variation in relaxation
rates as $T_{\mathrm{N}}$ is approached from below~\cite{Lancaster2004-CuX2(pyz)-and-0DMMs}.

\section{$\text{[Cu(pyo)}_{\text{6}}\text{]\emph{Z}}_{\text{2}}$\label{sec:[Cu(pyo)6]X2}}

\begin{figure}
\includegraphics{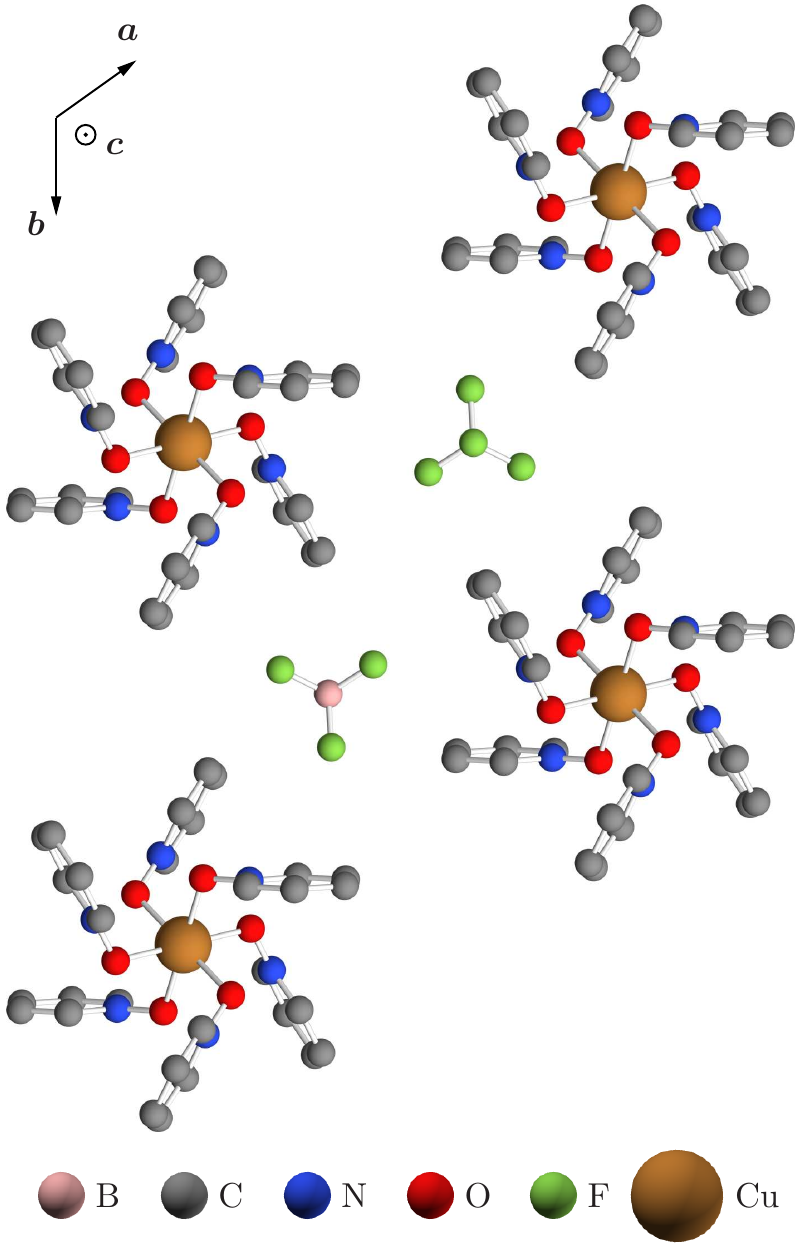} \caption{Structure of {[}Cu(pyo)$_{\text{6}}${]}(BF$_{\text{4}}$)$_{\text{2}}$,
viewed along the three-fold ($\bm{c}$-) axis, after Ref.~\onlinecite{Reinen1979-Cupyo6X2}.
Copper ions are surrounded by octahedra of six oxygens, each part
of a pyridine-\emph{N}-oxide ligand; {[}Cu(pyo)$_{\text{6}}${]}$^{\text{2+}}$
complexes space-pack with BF$_{\text{4}}^{-}$ stabilising the structure.
Ion sizes are schematic; copper ions are shown twice as large for
emphasis, and hydrogens have been omitted for clarity. As indicated
in the top left, the $\bm{a}$ and $\bm{b}$ directions lie in the
plane of the paper, separated by $\gamma=120^{\circ}$, whilst the
$\bm{c}$ direction is out of the page.\label{fig:Cu(pyo)6(BF4)2-structure}}
\end{figure}

The next example is not a coordination polymer, but instead forms
a three-dimensional structure of packed molecular groups. The molecular
magnet {[}Cu(pyo)$_{\text{6}}${]}\emph{Z}$_{\text{2}}$, where \emph{Z}$^{-}$~=
BF$_{\text{4}}^{-}$, ClO$_{\text{3}}^{-}$, PF$_{\text{6}}^{-}$,
comprises Cu$^{\text{2+}}$ ions on a slightly distorted cubic lattice,
located in {[}Cu(pyo)$_{\text{6}}${]}$^{\text{2+}}$ complexes, and
surrounded by octahedra of oxygen atoms~\cite{Algra1978-Cupyo6X2}.
The structure is shown in \ref{fig:Cu(pyo)6(BF4)2-structure}. This
approximately cubic structure, which arises from the molecules' packing,
might suggest that a three-dimensional model of magnetism would be
appropriate. In fact, although the observed bulk properties of {[}\emph{M}(pyo)$_{\text{6}}${]}\emph{X}$_{\text{2}}$
where \emph{M}$^{\text{2+}}$ ~= Co$^{\text{2+}}$, Ni$^{\text{2+}}$
or Fe$^{\text{2+}}$ are largely isotropic, but the copper analogues
display quasi--low-dimensional, $S=\frac{1}{2}$ Heisenberg antiferromagnetism~\cite{Algra1978-Cupyo6X2}.
Weakening of superexchange in certain directions, and thus the lowering
of the systems' effective dimensionality, is attributed to lengthening
of the superexchange pathways resulting from Jahn--Teller distortion
of the Cu--O octahedra, which is observed in structural and EPR measurements~\cite{Wood1980-Cupyo6X2,Reinen1979-Cupyo6X2}.
At high temperatures, ($T\gtrsim\mathrm{100\; K})$, these distortions
are expected to be dynamic but, as $T$ is reduced (to $\approx50\mathrm{\; K}$),
they freeze out. The anion \emph{Z}$^{-}$ determines the nature of
the static Jahn--Teller elongation. The \emph{Z}$^{-}$\emph{~}=
BF$_{\text{4}}^{-}$ material displays ferrodistortive ordering which,
in combination with the antiferromagnetic exchange, gives rise to
2D Heisenberg antiferromagnetic behavior~\cite{Algra1978-Cupyo6X2,Burriel1990-Cupyo6BF4-susceptibility}.
By contrast, \emph{Z}$^{-}$\emph{~}= ClO$_{\text{4}}^{-}$, NO$_{\text{3}}^{-}$
(neither of which is investigated here) display antiferrodistortive
ordering~\cite{Wood1980-Cupyo6X2}, which gives rise to quasi-1D
Heisenberg antiferromagnetism~\cite{Algra1978-Cupyo6X2}. All of
the samples investigated were measured in the LTF spectrometer at
S\murm S.

\begin{figure*}
\includegraphics{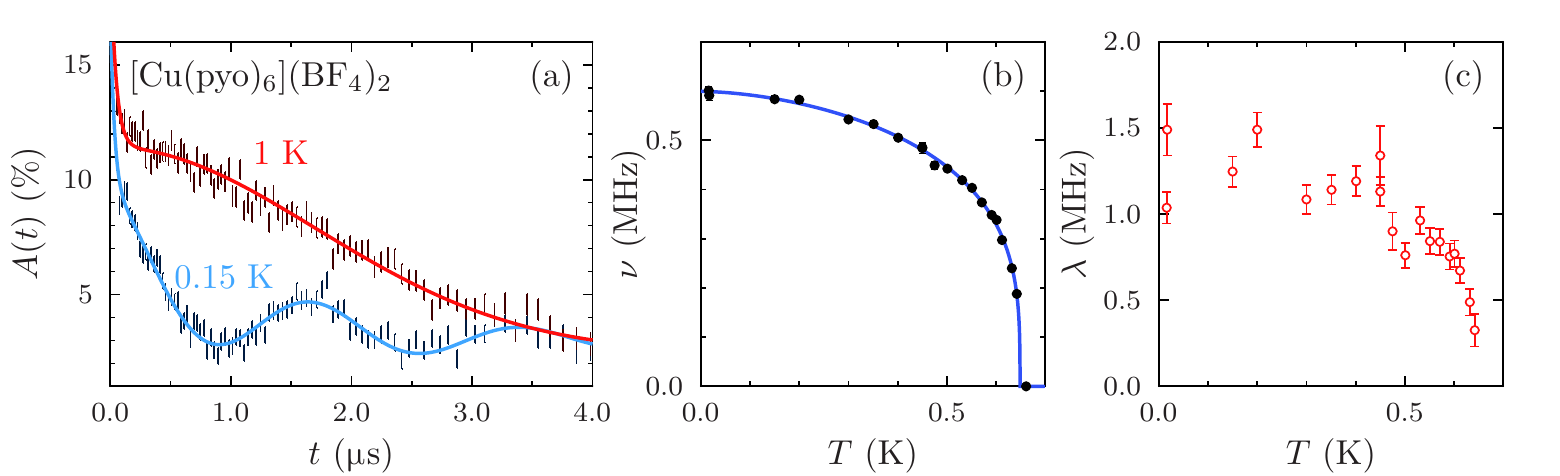}
\caption{Example data and fits for {[}Cu(pyo)$_{\text{6}}${]}(BF$_{\text{4}}$)$_{\text{2}}$.
From left to right: (a) shows sample asymmetry spectra $A(t)$ for
$T<T_{\mathrm{N}}$ and $T>T_{\mathrm{N}}$, along with fits to \prettyref{eq:Cu(pyz)2(pyo)2X2-A(t)}
and \prettyref{eq:Cu(pyo)6(BF4)2-highT}, respectively; (b) shows
frequency as a function of temperature; and (c) shows relaxation rate
$\lambda_{2}$ as a function of temperature; $\lambda_{1}$ was held
fixed during the fitting procedure, and $\lambda_{3}$ persists for
$T>T_{\mathrm{N}}$.  In the $\nu(T)$ plot, error bars are included
on the points but in most cases they are smaller than the marker being
used.\label{fig:Cu(pyo)6(BF4)2-magnetism-graphs}}
\end{figure*}

In the \emph{Z$^{-}$~}= BF$_{\text{4}}^{-}$ compound, below a temperature
$T_{\mathrm{N}}$, a single oscillating frequency is observed, indicating
a transition to a state of long-range magnetic order. Example data
above and below the transition, along with fits, are shown in \prettyref{fig:Cu(pyo)6(BF4)2-magnetism-graphs}~(a).
Data were fitted to \prettyref{eq:Cu(pyz)2(pyo)2X2-A(t)}, and the
frequencies extracted from the procedure fitted as a function of temperature
to \prettyref{eq:nuT1-T-over-TN-alpha-beta}as shown in \prettyref{fig:Cu(pyo)6(BF4)2-magnetism-graphs}~(b).
This procedure identifies a transition temperature $T_{\mathrm{N}}=0.649\pm0.005\mathrm{\; K}$.
The only other parameter found to vary significantly in the range
$20\mathrm{\; mK}\leq T\leq T_{\mathrm{N}}$ was $\lambda_{2}$, shown
in as shown in \prettyref{fig:Cu(pyo)6(BF4)2-magnetism-graphs}~(c).
The fitted parameters are shown in \prettyref{tab:other2dmm-params}.
We may compare this with the result of an earlier low-temperature
specific heat study~\cite{Algra1978-Cupyo6X2} which found a very
small $\lambda$-point anomaly at $T_{\mathrm{N}}=0.62\pm0.01\mathrm{\; K}$,
slightly lower than our result. Fitting the magnetic component of
the heat capacity with the predictions from a two-dimensional Heisenberg
antiferromagnet gives $J/k_{\mathrm{B}}=-1.10\pm0.02\mathrm{\; K}$,
and similar analysis of the magnetic susceptibility~\cite{Algra1978-Cupyo6X2}
yields $J/k_{\mathrm{B}}=-1.08\pm0.03\mathrm{\; K}$. Using these
values, together with the muon estimate of $T_{\mathrm{N}}$ and \prettyref{eq:Yasuda-J-Jprime},
allows us to estimate the inter-plane coupling, $J_{\perp}/J=0.26\pm0.01$.
(Using the value of $T_{\mathrm{N}}$ from heat capacity results in
an estimate $J_{\perp}/J=0.21\pm0.02$.)

For temperatures $T_{\mathrm{N}}<T\leq1\mathrm{\; K}$, the spectra
are well described by a relaxation function 
\begin{equation}
A(t)=A_{0}(p_{1}\mathrm{e}^{-\lambda t}+p_{2}\mathrm{e}^{-\sigma^{2}t^{2}})+A_{\mathrm{bg}},\label{eq:Cu(pyo)6(BF4)2-highT}
\end{equation}
 comprising an initial fast-relaxing component with $\lambda\approx20\mathrm{\mathrm{\; MHz}}$,
and a Gaussian relaxation with $\sigma\approx0.4\mathrm{\; MHz}$
corresponding to the slow depolarisation of muon spins due to randomly-orientated
nuclear moments.

\begin{figure*}
\includegraphics{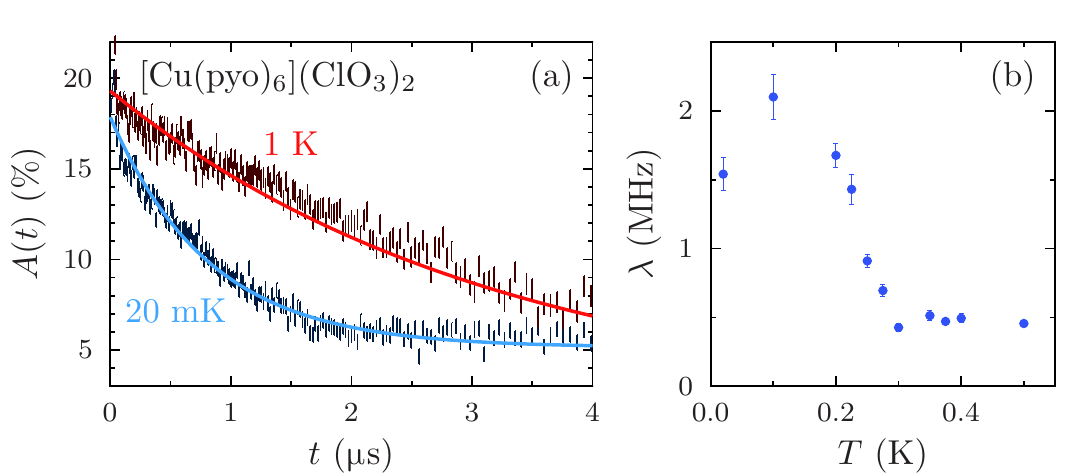}
\caption{Example data and fits for {[}Cu(pyo)$_{\text{6}}${]}(ClO$_{\text{3}}$)$_{\text{2}}$.
Representative $A(t)$ spectra for $T<T_{\mathrm{N}}$ and $T>T_{\mathrm{N}}$,
along with fits to \prettyref{eq:Cu(pyo)6(ClO3)2-A(t)}, are shown
in (a), whilst the value of the relaxation rate, $\lambda$, as a
function of temperature is shown in (b). \label{fig:Cupyo6ClO32}}
\end{figure*}

The \emph{Z$^{-}$~}= ClO$_{\text{3}}^{-}$ material also shows evidence
for a magnetic transition, although in this case we do not observe
oscillations in the muon asymmetry. Instead we measure a discontinuous
change in the relaxation which seems to point towards an ordering
transition. Example asymmetry spectra are shown in \prettyref{fig:Cupyo6ClO32}~(a)
and data at all measured temperatures are well described with the
relaxation function 
\begin{equation}
A(t)=A_{0}\mathrm{e}^{-\lambda t}+A_{\mathrm{bg}}\,.\label{eq:Cu(pyo)6(ClO3)2-A(t)}
\end{equation}
Evidence for a magnetic transition comes from the temperature evolution
of $\lambda$ {[}\prettyref{fig:Cupyo6ClO32}(b){]}, where we see
that the relaxation decreases with increasing temperature until it
settles at $T\approx0.3\mathrm{\; K}$, on a value $\lambda\approx0.5\mathrm{\; MHz}$.
It is likely that this tracks the internal magnetic field inside the
material, and is suggestive of $T_{\mathrm{N}}=0.30\pm0.01\mathrm{\; K}$.
\begin{figure}[b]
\includegraphics{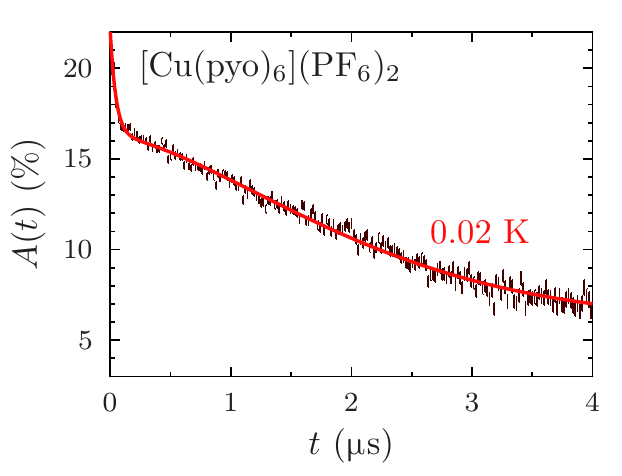}
\caption{Sample asymmetry spectrum for {[}Cu(pyo)$_{\text{6}}${]}(PF$_{\text{6}}$)$_{\text{2}}$
measured at $T=0.02\mathrm{\; K}$. The spectra remain indistinguishable
from this across the range of temperatures examined, $0.02\mathrm{\; K}\leq T\leq1\mathrm{\; K}$.
\label{fig:Cupyo6PF62}}
\end{figure}

The final member of this family studied, \emph{Z$^{-}$~}= PF$_{\text{6}}^{-}$,
shows no evidence for a magnetic transition over the range of temperatures
studied, $0.02\mathrm{\; K}\leq T\leq1\mathrm{\; K}$. An example
spectrum is shown in \prettyref{fig:Cupyo6PF62}. The data resemble
the above-transition data measured in the BF$_{\text{4}}$ and ClO$_{\text{3}}$
compounds and it therefore seems likely that the paramagnetic state
persists to the lowest temperature measured.

\section{$\text{Ag(pyz)}_{\text{2}}\text{(S}_{\text{2}}\text{O}_{\text{8}}\text{)}$\label{sec:Agpyz2S2O8}}

The examples so far have used Cu$^{\text{2+}}$ (3d$^{\text{9}}$)
as the magnetic species. An alternative strategy is to employ Ag$^{\text{2+}}$
(4d$^{\text{9}}$) which also carries an $S=\frac{1}{2}$ moment.
This idea has led to the synthesis of Ag(pyz)$_{\text{2}}$(S$_{\text{2}}$O$_{\text{8}}$),
which comprises square sheets of {[}Ag(pyz)$_{\text{2}}${]}$^{\text{2+}}$
units spaced with S$_{\text{2}}$O$_{\text{8}}$$^{\text{2}-}$ anions~\cite{Manson2009-Agpyz2S2O8}.
Each silver ion lies at the centre of an elongated (AgN$_{\text{4}}$O$_{\text{2}}$)
octahedron, where the Ag--N bonds are significantly shorter than the
Ag--O. Preparation details can be found in Ref.~\onlinecite{Matthews1971-Ag-Cu-Co-Ni-pyz-prep}.

\begin{figure*}
\includegraphics{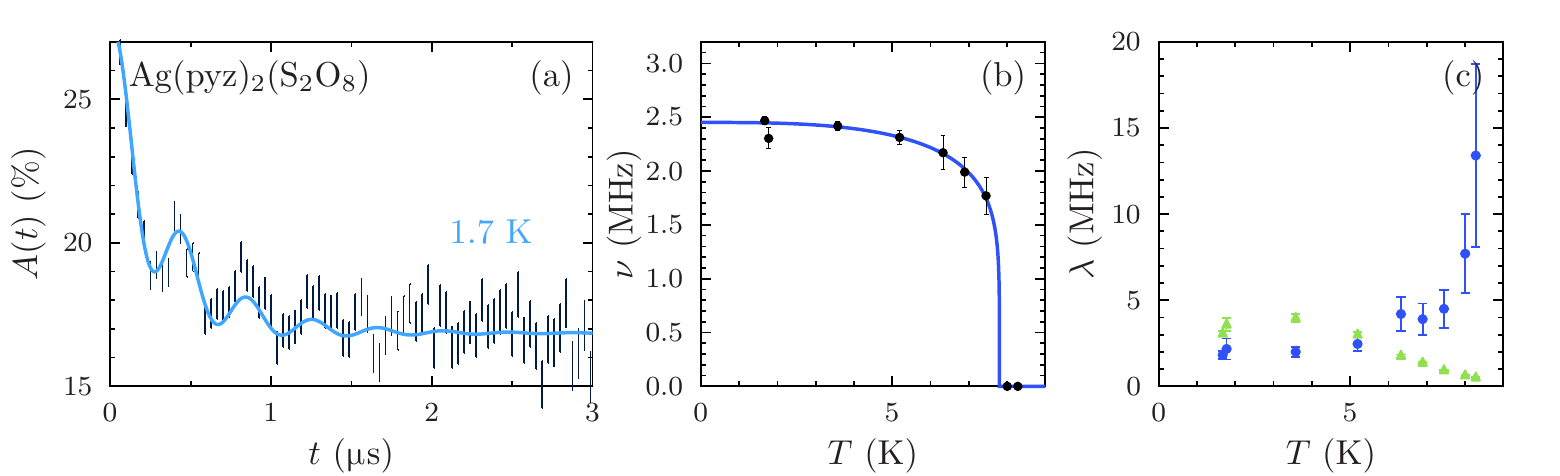}

\caption{Example data and fits for Ag(pyz)$_{\text{2}}$(S$_{\text{2}}$O$_{\text{8}}$).
From left to right: (a) shows sample asymmetry spectra $A(t)$ for
$T<T_{\mathrm{N}}$ along with a fit to \prettyref{eq:Ag(pyz)2(S2O8)-A(t)};
(b) shows the frequency as a function of temperature; and (c) shows
relaxation rates $\lambda_{i}$ as a function of temperature. Filled
circles show the relaxation rate $\lambda_{1}$, which relaxes the
oscillation. The filled triangles correspond to the relaxation $\lambda_{2}$.\label{fig:Ag2pyz2S2O8-magnetism-graphs}}
\end{figure*}
Measurements were made using the GPS instrument at S\murm S. Example
muon data, along with fits to various parameters, are shown in \prettyref{fig:Ag2pyz2S2O8-magnetism-graphs}.
Asymmetry oscillations are visible in spectra taken below a transition
temperature $T_{\mathrm{N}}$. The data were fitted with a relaxation
function
\begin{equation}
A(t)=A_{0}\left(p_{1}\cos(2\uppi\nu_{1}t+\phi_{1})\mathrm{e}^{-\lambda_{1}t}+p_{2}\mathrm{e}^{-\lambda_{2}t}\right)+A_{\mathrm{bg}},\label{eq:Ag(pyz)2(S2O8)-A(t)}
\end{equation}
comprising a single damped oscillatory component, a slow-relaxing
component, and a static background signal. The onset of increased
relaxation $\lambda_{1}$ as the transition is approached from below
leads to large statistical errors on fitted values, as is evident
in \prettyref{fig:Ag2pyz2S2O8-magnetism-graphs}~(c). The relaxation
$\lambda_{2}$ decreases with increasing temperature. Fitting to \prettyref{eq:nuT1-T-over-TN-alpha-beta}
allows the critical parameters $\beta=0.19\pm0.02$ and $T_{\mathrm{N}}=7.8\pm0.3\mathrm{\; K}$
to be determined. Fitted values are shown in \prettyref{tab:other2dmm-params}.

The in-plane exchange $J$ is too large to be determined with pulsed
fields~\cite{Manson2009-Agpyz2S2O8}: $M(B)$ does not saturate in
fields up to $64\mathrm{\; T}$. However, fitting $\chi(T)$ data
allows an estimate of the exchange $J/k_{\mathrm{B}}\approx53\mathrm{\; K}$
(and thus, in conjunction with $g$ measured by EPR, the saturation
field $B_{\mathrm{c}}$ is estimated to be $160\mathrm{\; T}$). Thus,
$k_{\mathrm{B}}T_{\mathrm{N}}/J=0.148\pm0.006$. Estimation of the
exhange anisotropy with \prettyref{eq:Yasuda-J-Jprime} yields $|J_{\perp}/J|\sim10^{-6}$,
but this very small ratio of ordering temperature to exchange strength
is outside the range in which the equation is known to yield accurate
results. The alternative method of parametrizing the low dimensionality
in terms of correlation length at the Néel temperature (see \prettyref{sub:exchange-anisotropy})
yields $\xi(T_{\mathrm{N}})/a=1000\pm300$.

\section{$\text{[Ni(HF}_{\text{2}}\text{)(pyz)}_{\text{2}}\text{]\emph{X}}$\label{sec:dog}}

\begin{table*}
\centering{}%
\begin{tabular}{cccccccccc}
\toprule 
material & $\nu_{1}\mathrm{\,(MHz)}$  & $p_{1}$  & $p_{2}$  & $p_{3}$  & $T_{\mathrm{N}}\mathrm{\,(K)}$  & $\beta$  & $\alpha$  & $J/k_{\mathrm{B}}\mathrm{\:(K)}$  & $|J_{\perp}/J|$\tabularnewline
\midrule 
{[}Cu(pyz)$_{\text{2}}$(pyo)$_{\text{2}}${]}(BF$_{\text{4}}$)$_{\text{2}}$ & $1.4(1)$  & $<10$  & $50$  & $50$  & $1.5(1)$  & $0.25(10)$  & $1.6(3)$  & -  & -\tabularnewline
{[}Cu(pyz)$_{\text{2}}$(pyo)$_{\text{2}}${]}(PF$_{\text{6}}$)$_{\text{2}}$ & $0.64(1)$  & $<10$  & $50$  & $50$  & $1.72(2)$  & $0.22(2)$  & $1.1(3)$  & $8.1(3)$ {*} & $2\times10^{-4}$\tabularnewline
{[}Cu(pyo)$_{\text{6}}${]}(BF$_{\text{4}}$)$_{\text{2}}$ & $0.599(5)$  & $15$  & $30$  & $55$  & $0.649(5)$  & $0.29(1)$  & $1.7(1)$  & $1.09(2)$ $^{\dagger}$ & $0.23(2)$\tabularnewline
Ag(pyz)$_{\text{2}}$(S$_{\text{2}}$O$_{\text{8}}$) & $2.45(4)$ & $30$ & $70$ & - & $7.8(3)$ & $0.19(2)$ & $3(2)$ & $52.7(3)$ & $\sim10^{-6}$\tabularnewline
\bottomrule
\end{tabular}\caption{Fitted parameters for molecular magnets in the family {[}Cu(pyz)$_{\text{2}}$(pyo)$_{\text{2}}${]}\emph{Y}$_{\text{2}}$,
{[}Cu(pyo)\textsubscript{6}{]}(BF\textsubscript{4})\textsubscript{2}
and Ag(pyz)$_{\text{2}}$(S$_{\text{2}}$O$_{\text{8}}$). The first
parameters shown relate to fits to \prettyref{eq:Cu(pyz)2(pyo)2X2-A(t)}
(for the first three rows) or \prettyref{eq:Ag(pyz)2(S2O8)-A(t)}
{[}for Ag(pyz)$_{\text{2}}$(S$_{\text{2}}$O$_{\text{8}}$){]}. This
allows us to derive frequencies at $T=0$, $\nu_{i}$; and probabilities
of stopping in the various classes of stopping site, $p_{i}$, in
percent. Then, the temperature dependence of $\nu_{i}$ is fitted
with \prettyref{eq:nuT1-T-over-TN-alpha-beta}, extracting values
for the Néel temperature, $T_{\mathrm{N}}$, critical exponent $\beta$
and parameter $\alpha$. Finally, the quoted $J/k_{\mathrm{B}}$ is
obtained from pulsed-field experiments~\cite{goddard2008-2DHMexchange}
The asterisk ({*}) indicates the value of $J$ was obtained using
$g_{ab}=2.04(1)$ in conjunction with single-crystal pulsed-field
data. The dagger ($^{\dagger}$) indicates that the value of $J$
is extracted from heat capacity and susceptibility from Ref.~\onlinecite{Algra1978-Cupyo6X2}.).
The ratio of inter- to in-plane coupling, $J_{\perp}/J$, is obtained
by combining $T_{\mathrm{N}}$ and $J$ with formulae extracted from
quantum Monte Carlo simulations (see \prettyref{sub:exchange-anisotropy},
and Ref.~\onlinecite{goddard2008-2DHMexchange}). The dash in the
$p_{3}$ column for Ag(pyz)$_{\text{2}}$(S$_{\text{2}}$O$_{\text{8}}$)
reflects the fact that there is no third component in \prettyref{eq:Ag(pyz)2(S2O8)-A(t)}.
Dashes in the $J/k_{\mathrm{B}}$ and $J_{\perp}/J$ columns for {[}Cu(pyz)$_{\text{2}}$(pyo)$_{\text{2}}${]}(BF$_{\text{4}}$)$_{\text{2}}$
indicate a lack of pulsed-field data for this material.\label{tab:other2dmm-params}}
\end{table*}

In order to investigate the influence of a different spin state on
the magnetic cation in the {[}\emph{M}(HF$_{\text{2}}$)(pyz)$_{\text{2}}${]}\emph{X}
architecture the {[}Ni(HF$_{\text{2}}$)(pyz)$_{\text{2}}${]}\emph{X}
(\emph{X$^{-}$~}= PF$_{\text{6}}^{-}$, SbF$_{\text{6}}^{-}$) system
has been synthesised~\cite{Manson2011-NiHF2pyz2X}. These materials
are isostructural with the copper family discussed in \prettyref{sec:[Cu(HF2)(pyz)2]X},
but contain $S=1$ Ni$^{\text{2+}}$ cations.

\begin{figure*}
\includegraphics{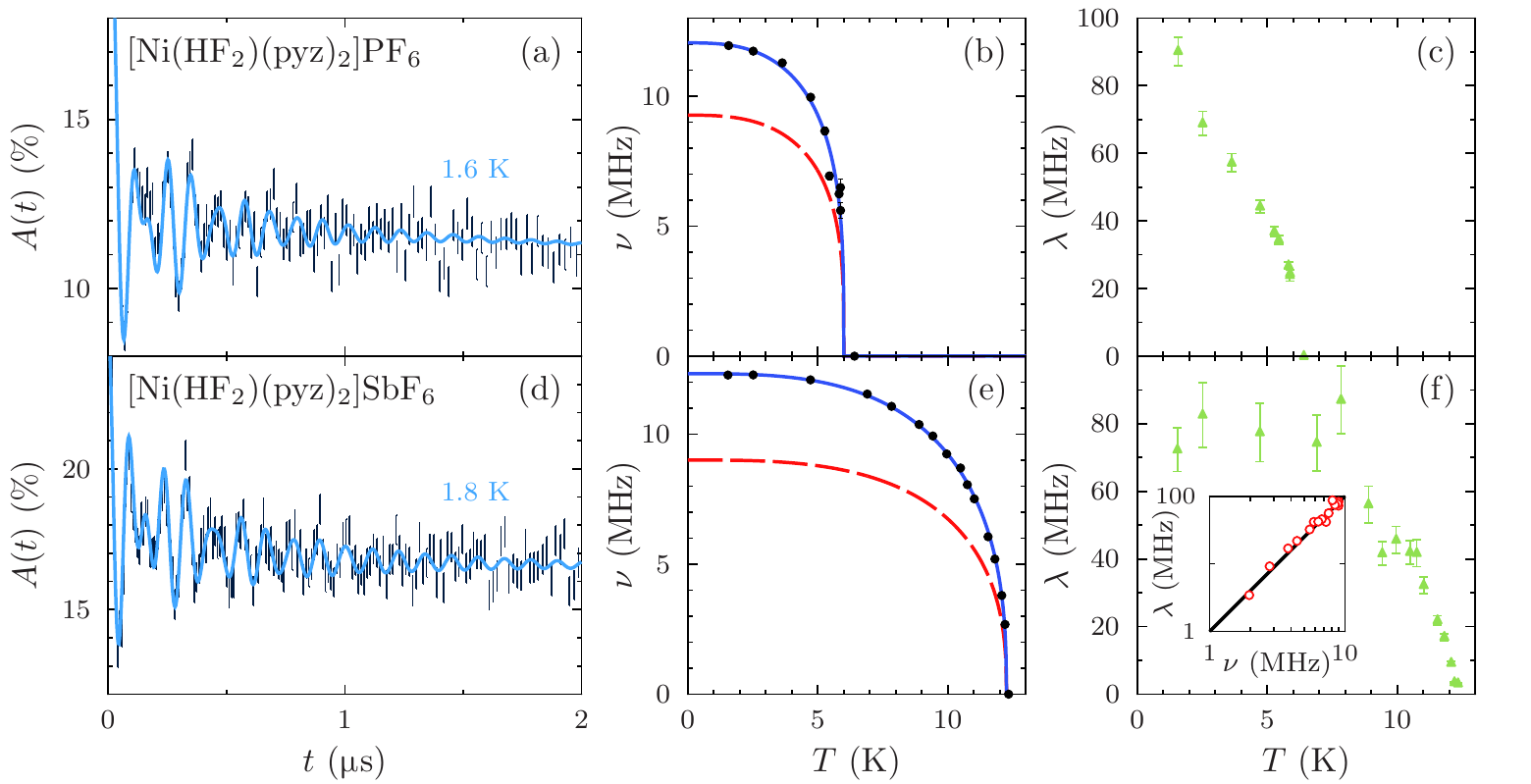}
\caption{Example data and fits for \emph{M~}= Ni magnets. From left to right:
(a) and (d) show sample asymmetry spectra $A(t)$ for $T<T_{\mathrm{N}}$
along with a fit to \prettyref{eq:M(HF2)(pyz)2X-A(t)-fit}; (b) and
(e) show frequencies as a function of temperature {[}no data points
are shown for the second line because this frequency $\nu_{2}$ was
held in fixed proportion to the first, $\nu_{1}$ (see text){]}; and
(c) and (f) show relaxation rate $\lambda_{3}$ as a function of temperature.
In the $\nu(T)$ plot, error bars are included on the points but in
most cases they are smaller than the marker being used. The inset
to (f) shows $\nu_{2}\equiv\nu$ plotted against $\lambda_{3}\equiv\lambda$
on a log--log scale. The black line shows $\lambda=\nu^{2}$ .\label{fig:Ni(HF2)(pyz)2X-magnetism-graphs}}
\end{figure*}

\begin{table*}
\begin{tabular}{cccccccccc}
\toprule 
\emph{X}  & $T_{\mathrm{N}}\mathrm{\:(K)}$  & $\nu_{1}\mathrm{\:(MHz)}$  & $\nu_{2}\mathrm{\:(MHz)}$  & $\lambda_{3}\mathrm{\;(MHz)}$  & $p_{1}$  & $p_{2}$  & $p_{3}$  & $\beta$  & $\alpha$ \tabularnewline
\midrule 
PF$_{\text{6}}$  & $6.0(4)$  & $12.0$  & $9.3$  & $90$  & $15$  & $10$  & $75$  & $0.25(10)$  & $2.8(2)$ \tabularnewline
SbF$_{\text{6}}$  & $12.26(1)$  & $12.3(1)$  & $8.98(1)$ & $80$  & $25$  & $10$  & $65$  & $0.34(4)$  & $3.1(1)$ \tabularnewline
\bottomrule
\end{tabular}\caption{Fitted parameters for \emph{M~}= Ni magnets. Errors shown are statistical
uncertainties on fitting and thus represent lower bounds. Errors on
Ni...PF$_{\text{6}}$ could not be estimated due to the fitting procedure
(see text).\label{tab:Ni(HF2)(pyz)2X-params}}
\end{table*}

Data were taken using the GPS spectrometer at PSI. Example data are
shown in Fig.~\ref{fig:Ni(HF2)(pyz)2X-magnetism-graphs}. We observe
oscillations at two frequencies below the materials' respective ordering
temperatures. Data were fitted with a relaxation function 
\begin{eqnarray}
A(t) & = & A_{0}\left[p_{1}\mathrm{e}^{-\lambda_{1}t}\cos(2\pi\nu_{1}t)+p_{2}\mathrm{e}^{-\lambda_{2}t}\cos(2\pi P_{2}\nu_{1}t)\right.\nonumber \\
 &  & \left.+p_{3}\mathrm{e}^{-\lambda_{3}t}\right]+A_{\mathrm{bg}}\mathrm{e}^{-\lambda_{\mathrm{bg}}t}.\label{eq:Ni(HF2)(pyz)2X-A(t)-fit}
\end{eqnarray}
 %where $A_{0}$ represents those muons which stop inside the sample
%and $A_{\mathrm{bg}}$ accounts for a relaxing background signal due
%to those muons that stop in the silver sample holder or cryostat tails,
%and the $\frac{1}{3}$ component of muon-spin which is parallel to
%the internal field. 
Of those muons which stop in the sample, $p_{1}\approx25\%$ indicates
the fraction of the signal corresponding to the low-frequency oscillating
state with $\nu_{1}(T=0)\approx12.3\mathrm{\: MHz}$; $p_{2}\approx10\%$
corresponds to muons stopping in the high-frequency oscillating state
with $\nu_{2}(T=0)\approx9.0\mathrm{\; MHz}$; and $p_{3}\approx65\%$
represents muons stopping in a site with a large relaxation rate $\lambda_{3}(T=0)\approx70\mathrm{\; MHz}$.
The frequencies were observed to scale with one-another, and consequently
the second frequency was held in fixed proportion $\nu_{2}=P_{2}\nu_{1}$
during the fitting procedure. The only other parameter which changes
significantly in value below $T_{\mathrm{N}}$ is $\lambda_{3}$,
which decreases with a trend qualitatively similar to that of the
frequencies. Fitting the extracted frequencies to \prettyref{eq:nuT1-T-over-TN-alpha-beta}
allows the transition temperature $T_{\mathrm{N}}=12.25\pm0.03\mathrm{\: K}$
and critical exponent $\beta=0.34\pm0.04$ to be extracted. In contrast
to the copper family studied in \prettyref{sec:[Cu(HF2)(pyz)2]X},
the relation $\lambda\propto\nu^{2}$ holds true, suggesting that
a field distribution whose width diminishes with increasing temperature
is responsible for the variation in $\lambda$, and that dynamics
are relatively unimportant in determining the muon response. This
is shown graphically in the inset to \prettyref{fig:Ni(HF2)(pyz)2X-magnetism-graphs}~(f),
where a plot of frequency against relaxation rate lies on top of a
line representing a $\lambda=\nu^{2}$ relationship. The phase $\phi$
required in previous fits (e.g. \prettyref{eq:M(HF2)(pyz)2X-A(t)-fit})
is not necessesary in fitting these spectra, and is set to zero. Fitted
parameters are shown in \prettyref{tab:Ni(HF2)(pyz)2X-params}.

Data for the \emph{X}$^{-}$~= PF$_{\text{6}}^{-}$ compound was
subject to similar analysis, fitting spectra below $T_{\mathrm{N}}$
to \prettyref{eq:Ni(HF2)(pyz)2X-A(t)-fit}, this time with $p_{1}\approx15\%$,
$\nu_{1}(T=0)\approx12.0\mathrm{\: MHz}$; $p_{2}\approx10\%$, $\nu_{2}(T=0)\approx9.3\mathrm{\: MHz}$;
and $p_{3}\approx75\%$, $\lambda_{3}(T=0)\approx100\mathrm{\; MHz}$.
The phase $\phi$ again proved unnecessary. These spectra do not show
as sharp a transition as the \emph{X}$^{-}$~= SbF$_{\text{6}}^{-}$
compound, with the oscillating fraction of the signal decaying rather
before the appearance of spectra whose different character indicates
clearly that the sample is above $T_{\mathrm{N}}$. The available
data do not allow reliable extraction of critical parameters, but
we estimate $5.5\mathrm{\; K}\leq T_{\mathrm{N}}\leq6.2\mathrm{\; K}$
and $0.15\leq\beta\leq0.4$. Fitted values are shown in \prettyref{tab:Ni(HF2)(pyz)2X-params}.

\begin{figure}
\includegraphics{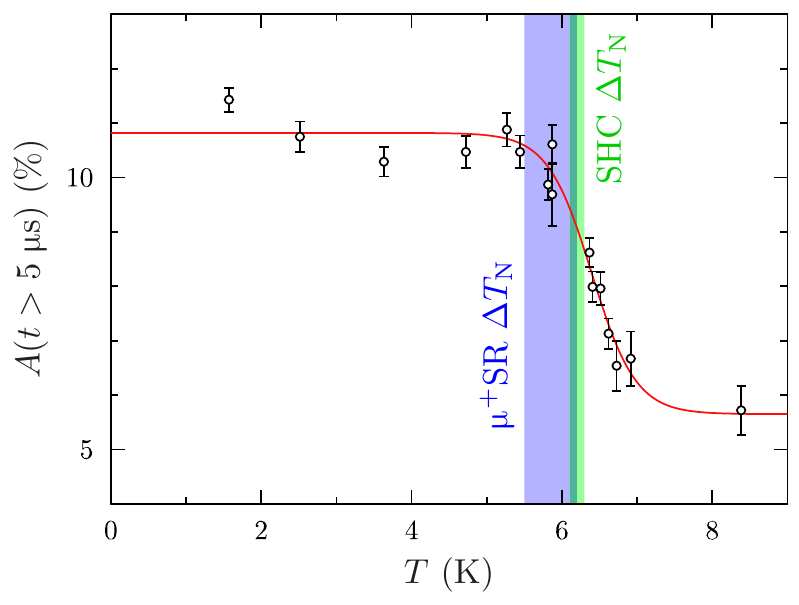}
\caption{The asymmetry at late times $A(t>5\mathrm{\;\murm s)}$ measured in
Ni(HF$_{\text{2}}$)(pyz)$_{\text{2}}$PF$_{\text{6}}$ along with
a fit to \prettyref{eq:NiPF6-Alatet-Fermi-step}. Overlaid are regions
corresponding to the range of values found by fitting the frequency
of oscillations as a function of temperature with \prettyref{eq:nuT1-T-over-TN-alpha-beta}
($5.5\mathrm{\; K}\leq T_{\mathrm{N}}\leq6.2\mathrm{\; K}$), and
the value of the $\lambda$-like anomaly in specific heat capacity
(SHC) measurements with a representative error of $0.1\mathrm{\; K}$
($6.1\mathrm{\; K}\leq T_{\mathrm{N}}\leq6.3\mathrm{\; K}$). The
darker area represents the overlap between these regions.\label{fig:NiPF6-Alatet}}
\end{figure}

Another method to locate the transition is to observe a transition
in the amplitude of the muon spectra at late times to observe the
transition as a function of temperature from zero in the unordered
state to the `$\frac{1}{3}$-tail' characteristic of LRO, described
in \prettyref{sec:Experimental-details}. Spectra were fitted with
the simple relaxation function $A(t>5\mathrm{\;\murm s})=A_{\mathrm{bg}}\mathrm{e}^{-\lambda_{\mathrm{bg}}t}$,
and then the amplitudes obtained fitted with a Fermi-like step function
\begin{equation}
A(t>5\mathrm{\;\murm s},\; T)=A_{2}+\frac{A_{1}-A_{2}}{\mathrm{e}^{(T-T_{\mathrm{mid}})/w}+1}\,,\label{eq:NiPF6-Alatet-Fermi-step}
\end{equation}
which provides a method of modelling a smooth transition between $A_{1}=A(T<T_{\mathrm{N}})$
and $A_{2}=A(T>T_{\mathrm{N}})$. The fitted amplitudes and Fermi
function are shown in \prettyref{fig:NiPF6-Alatet} (c). The fitted
mid-point $T_{\mathrm{mid}}=6.4\pm0.1\mathrm{\; K}$, and width $w=0.3\pm0.1\mathrm{\; K}$.
Spin relaxation peaks just above $T_{\mathrm{N}}$, and so one would
expect that $T_{\mathrm{N}}$ lies at the lower end of this transition.
Thus, the \muSR\ analysis suggests $T_{\mathrm{N}}=6.1\pm0.3\mathrm{\; K}$
(i.e.~$T_{\mathrm{mid}}-w\pm w$). This is consistent with the estimate
from $\nu_{i}(T)$ and the value $T_{\mathrm{N}}=6.2\mathrm{\; K}$
obtained from heat capacity~\cite{Manson2011-NiHF2pyz2X}.

Members of the \emph{M}~= Ni family exhibit \Fmu\ oscillations
rather like their copper counterparts, with a similar fraction of
the muons in sites giving rise to dipole--dipole interactions. The
results of these fits are shown along with those from Cu compounds
in \prettyref{tab:Cu(HF2)(pyz)2X-muF-parameters}. Because the nickel
data were measured at temperatures different from those of the copper
compounds and, as described in \prettyref{sub:[Cu(HF2)(pyz)2]X-paramagnetic},
the muon--fluorine bond length in the compound is sensitive to changes
in temperature, care must be taken when comparing these values to
those of the Cu family. Linearly interpolating the bond lengths for
{[}Cu(HF$_{\text{2}}$)(pyz)$_{\text{2}}${]}SbF$_{\text{6}}$ at
$9\mathrm{\; K}$ and $29\mathrm{\; K}$ to find an approximate value
of the bond length at $19\mathrm{\; K}$ yields $r_{\text{\murm--F}}(T=19\mathrm{\; K})=0.1062\pm0.0002\mathrm{\; nm}$
(where the error represents a combination of the statistical errors
on the fits and variation recorded in thermometry, and is thus a lower
bound), very similar to that measured for {[}Ni(HF$_{\text{2}}$)(pyz)$_{\text{2}}${]}SbF$_{\text{6}}$,
$r_{\text{\murm--F}}(T=19\mathrm{\; K})=0.1068\pm0.0004\mathrm{\; nm}$.

In spite of being isostructural to the {[}Cu(HF$_{\text{2}}$)(pyz)$_{\text{2}}${]}\emph{X}
systems, the dimensionality of these Ni variants is ambiguous. Susceptibility
data fit acceptably to a number of models, and \emph{ab initio} theoretical
calculations are suggestive of one-dimensional behavior, dominated
by the exchange along the bifluoride bridges. This is discussed more
fully in Ref.~\onlinecite{Manson2011-NiHF2pyz2X}.

\section{Discussion}

\begin{figure}
\includegraphics{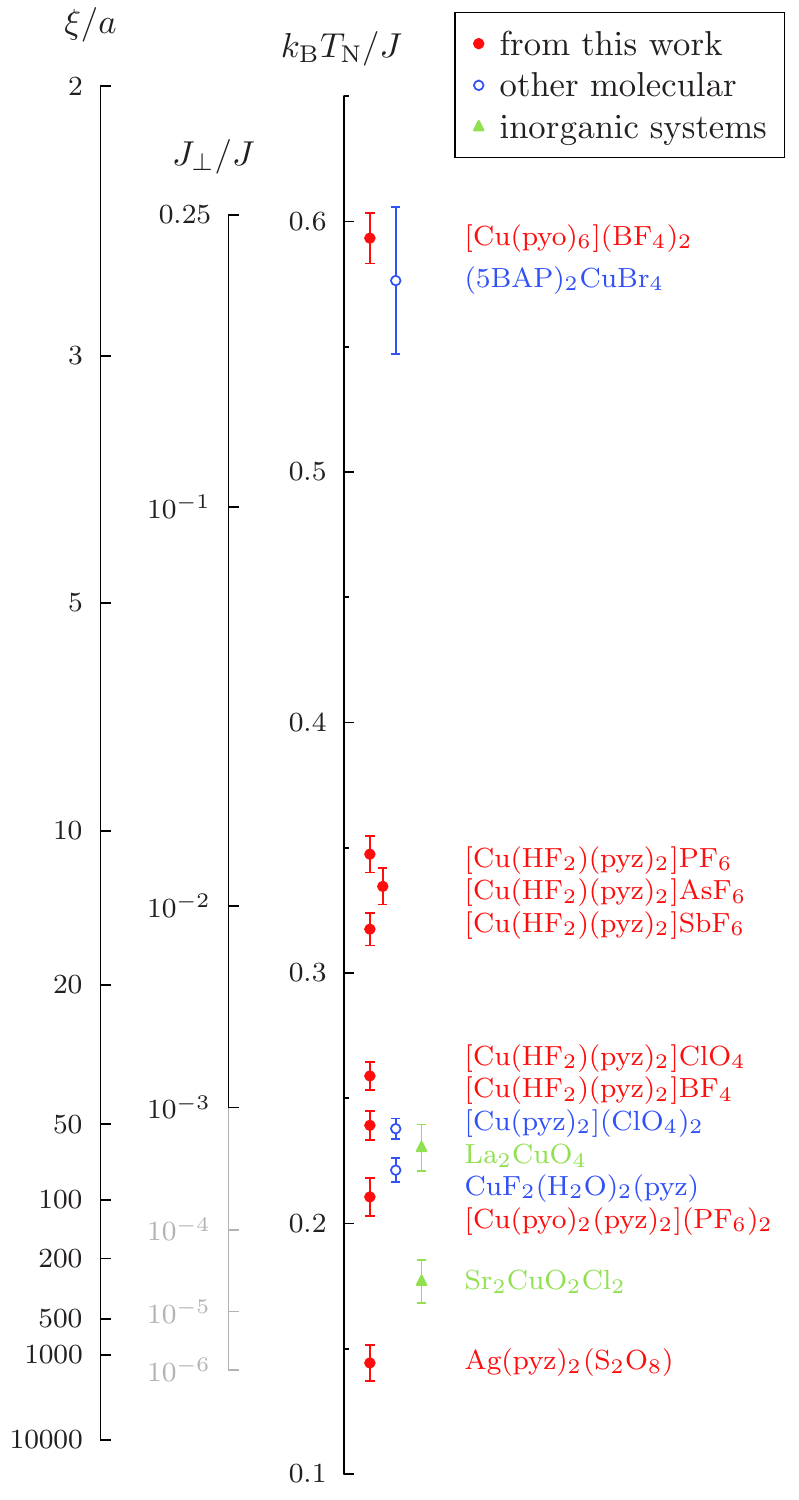}\caption{Quantification of the two-dimensionality of the  materials in this
paper and comparison with other notable 2DSLQHA systems. Filled circles
show materials investigated in this paper; open circles show other
molecular materials; and filled triangles show inorganic systems.
The low dimensionality is parametrized firstly with the directly experimental
ratio $T_{\mathrm{N}}/J$, and then with predictions quantum Monte
Carlo simulations for both the exchange anisotropy, $J_{\perp}/J$
{[}\prettyref{eq:Yasuda-J-Jprime}{]}, and the correlation length
of an ideal 2D Heisenberg antiferromagnet with the measured $J$ at
a temperature $T_{\mathrm{N}}$ {[}\prettyref{eq:Beard-correlation-length}{]}.
The greying-out of the axis for $J_{\perp}/J$ indicates where \prettyref{eq:Yasuda-J-Jprime}
is extrapolated beyond the range for which it was originally derived~\cite{yasuda2005-2DHA-JvsTN}.
Values of $T_{\mathrm{N}}/J$ for the other materials were evaluated
from Refs.~\onlinecite{Woodward2001-5BAP2CuBr4,Greven1995-Sr2CuO2Cl2,Aeppli1989-La2CuO4-Ba-doping}.\label{fig:exchange-anisotropy-review}}
\end{figure}

%Molecular chemistry appears to be a promising avenue for synthesising
%novel low-dimensional model systems. 
\prettyref{fig:exchange-anisotropy-review} collects the results from
this paper and shows how isolation between two-dimensional layers
varies over a variety of systems; those presented in this paper, molecular
materials studied elsewhere, and inorganic materials. The primary
axis is the experimental ratio $T_{\mathrm{N}}/J$. Also shown are
the ratios of the in-plane and inter-plane exchange interactions,
$J_{\perp}/J$, and the correlation length at the transition, $\xi(T_{\mathrm{N}})/a$,
extracted from fits to quantum Monte Carlo simulations. Of the materials
in this paper, the least anisotropic is {[}Cu(pyo)$_{\text{6}}${]}(BF$_{\text{4}}$)$_{\text{2}}$,
whose low transition temperature is caused by a small exchange constant
rather than particularly high exchange anisotropy. The members of
the {[}Cu(HF$_{\text{2}}$)(pyz)$_{\text{2}}${]}\emph{X} family with
octahedral anions have $k_{\mathrm{B}}T_{\mathrm{N}}/J\approx0.33$,
rather more than the $k_{\mathrm{B}}T_{\mathrm{N}}/J\approx0.25$
shown by their counterparts with tetrahedral anions (as has been noted
previously~\cite{goddard2008-2DHMexchange}). This makes the latter
comparable to highly 2D molecular systems {[}Cu(pyz)$_{\text{2}}${]}(ClO$_{\text{4}}$)$_{\text{2}}$
and CuF$_{\text{2}}$(H$_{\text{2}}$O)$_{\text{2}}$(pyz), and the
cuprate parent compound La$_{\text{2}}$CuO$_{\text{4}}$. Below this,
{[}Cu(pyo)$_{\text{2}}$(pyz)$_{\text{2}}${]}(PF$_{\text{6}}$)$_{\text{2}}$
exhibits a ratio $k_{\mathrm{B}}T_{\mathrm{N}}/J\approx0.21(1)$.
The prototypical inorganic 2D system Sr$_{\text{2}}$CuO$_{\text{2}}$Cl$_{\text{2}}$
exhibits $k_{\mathrm{B}}T_{\mathrm{N}}/J=0.177\pm0.009$, and thus
$\xi(T_{\mathrm{N}})/a=280\pm90$. The most 2D material investigated
in this paper, Ag(pyz)$_{\text{2}}$(S$_{\text{2}}$O$_{\text{8}}$),
has $k_{\mathrm{B}}T_{\mathrm{N}}/J=0.148\pm0.006$, implying $\xi(T_{\mathrm{N}})/a=1000\pm300$,
with the added benefit that the magnetic field required to probe its
interactions is far closer to the range of fields achievable in the
laboratory. By these measures, molecular magnets provide some excellent
realizations of the 2DSLQHA, with Ag(pyz)$_{\text{2}}$(S$_{\text{2}}$O$_{\text{8}}$)
being the best realization found to date.

Another method of examining the dimensionality of these systems is
to consider their behavior in the critical region. The critical exponent
$\beta$ is a quantity frequently extracted in studies of magnetic
materials, and it is often used to make inferences about the dimensionality
of the system under study. In the critical region near a magnetic
transition, an order parameter $\Phi$, identical to the (staggered)
magnetization, would be expected to vary as
\begin{equation}
\Phi(T)=\Phi_{0}\left(1-\frac{T}{T_{\mathrm{N}}}\right)^{\beta}.\label{eq:critical-beta}
\end{equation}
In simple, isotropic cases, the value of $\beta$ depends on the
dimensionality of the system, $d$, and that of the order parameter,
$D$. For example, in the 3D Heisenberg model ($d=3$, $D=3$), $\beta=0.367$,
whilst in the 2D Ising model ($d=2$, $D=1$), $\beta=\frac{1}{8}$.
Since the muon precession frequency is proportional to the local field,
it is also proportional to the moment on the magnetic ions in a crystal,
and can be used as an effective order parameter. However, \prettyref{eq:critical-beta}
would only be expected to hold true in the critical region. The extent
of the critical region (defined as that region where simple mean-field
theory does not apply) can be parametrized by the Ginzburg temperature
$T_{\mathrm{G}}$, which is related to the transition temperature
$T_{\mathrm{c}}$ by~\cite{Chaikin-and-Lubensky}
\begin{equation}
\frac{|T_{\mathrm{G}}-T_{\mathrm{c}}|}{T_{\mathrm{c}}}=\left[\left(\frac{\xi}{a}\right)^{d}\left(\frac{\Delta C}{k_{\mathrm{B}}}\right)\right]^{\frac{2}{d-4}},
\end{equation}
where $d$ is the dimensionality, $\xi$ is the correlation length
and $\Delta C$ is the discontinuity in the heat capacity. Quantum
Monte Carlo simulations suggest~\cite{sengupta2003-bulkmeasuresbad}
that $\Delta C/k_{\mathrm{B}}\approx J_{\perp}/J$. It follows that
for $d=3$, where $\Delta C/k_{\mathrm{B}}\approx1$ we have $\left|T_{\mathrm{G}}-T_{\mathrm{c}}\right|/T_{\mathrm{c}}\approx(\xi/a)^{-6}$,
giving rise to a narrow critical region. In two dimensions, we have
$\left|T_{\mathrm{G}}-T_{\mathrm{c}}\right|/T_{\mathrm{c}}\approx(\xi/a)^{-2}(\Delta C/k_{\mathrm{B}})^{-1}$.
Anisotropic materials with small $J_{\perp}/J$ only show a small
heat capacity discontinuity, while $\xi/a$ grows according to \prettyref{eq:Beard-correlation-length}.
This leads to a $\left|T_{\mathrm{G}}-T_{\mathrm{c}}\right|/T_{\mathrm{c}}$
of order 1 for our materials, that is, a larger critical region for
2D (as compared to 3D) systems.

\begin{figure}
\includegraphics{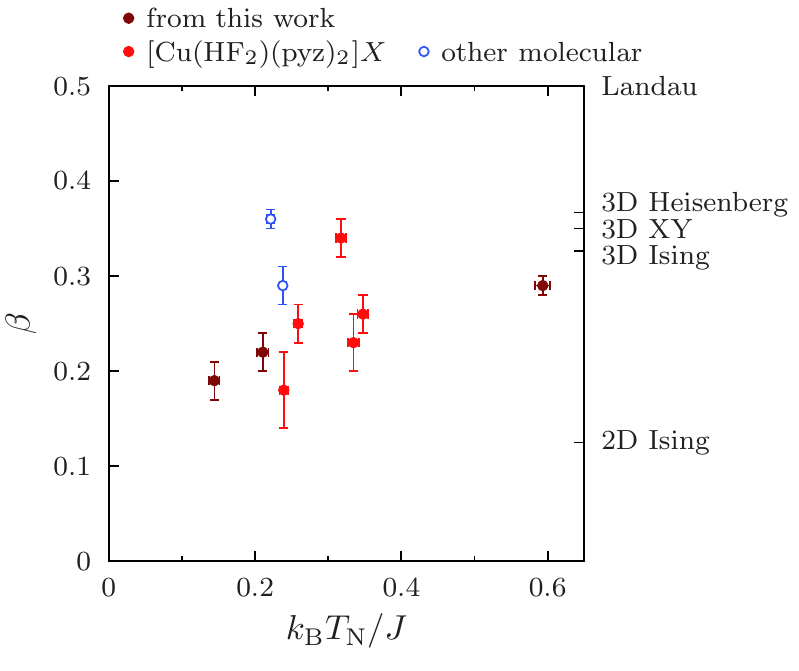} \caption{Critical exponent $\beta$, as commonly extracted from \prettyref{eq:nuT1-T-over-TN-alpha-beta},
plotted against the experimental ratio $k_{\mathrm{B}}T_{\mathrm{N}}/J$,
indicative of exchange anisotropy. Bright filled circles indicate
{[}Cu(HF$_{\text{2}}$)(pyz)$_{\text{2}}${]}\emph{X} materials examined
in \prettyref{sec:[Cu(HF2)(pyz)2]X}, whilst darker circles indicate
other materials studied in this work: {[}Cu(pyz)$_{\text{2}}$(pyo)$_{\text{2}}${]}BF$_{\text{4}}$
(\prettyref{sec:[Cu(pyz)2(pyo)2]X2}); {[}Cu(pyo)$_{\text{6}}${]}(BF$_{\text{4}}$)$_{\text{2}}$
(\prettyref{sec:[Cu(pyo)6]X2}); Ag(pyz)$_{\text{2}}$(S$_{\text{2}}$O$_{\text{8}}$)
(\prettyref{sec:Agpyz2S2O8}). Open circles indicate other molecular
materials: CuF$_{\text{2}}$(H$_{\text{2}}$O)$_{\text{2}}$(pyz)
(Ref.~\onlinecite{Manson2008-CuF2(H2O)2(pyz)}); {[}Cu(pyz)$_{\text{2}}${]}(ClO$_{\text{4}}$)$_{\text{2}}$
(Ref.~\onlinecite{lancaster2007-CuPz2ClO42}).\label{fig:beta-vs-TNJ}}
\end{figure}

The large critical region in these materials allows meaningful critical
parameters to be extracted from muon data. The simplest method of
doing so is to fit the data to \prettyref{eq:nuT1-T-over-TN-alpha-beta},
as we have throughout this study; alternatively, critical scaling
plots can be used (e.g. Ref.~\onlinecite{Pratt2007-critical-behaviour-CoGly}),
which we have performed, finding the results are unchanged within
error. Since $\beta$ might be expected to give an indication of the
dimensionality of the hydrodynamical fluctuations in these materials,
a comparison between extracted $\beta$ and exchange anisotropy parametrized
by $k_{\mathrm{B}}T_{\mathrm{N}}/J$ is shown in \prettyref{fig:beta-vs-TNJ}.
Members of the {[}Cu(HF$_{\text{2}}$)(pyz)$_{\text{2}}${]}\emph{X}
family show some correlation between the critical exponent and the
effective dimensionality but overall, the relationship is weak. This
is probably because $\beta$, which is not a Hamiltonian parameter,
is not simply a function of the dimensionality of the interactions,
but probes the nature of the critical dynamics (including propagating
and diffusive modes) which could differ substantially between systems.

\section{Conclusions}

We have presented a systematic study of muon-spin relaxation measurements
on several families of quasi two-dimensional molecular antiferromagnet,
comprising ligands of pyrazine, bifluoride and pyridine-\emph{N}-oxide;
and the magnetic metal cations Cu$^{\text{2+}}$, Ag$^{\text{2+}}$
and Ni$^{\text{2+}}$. In each case \muSR\  has been shown to be
sensitive to the transition temperature $T_{\mathrm{N}}$, which is
often difficult to unambiguously identify with specific heat and magnetic
susceptibility measurements. We have combined these measurements with
predictions of quantum Monte Carlo calculations to identify the extent
to which each is a good realization of the 2DSLQHA model. The critical
parameters derived from following the temperature evolution of the
\muSR\  precession frequencies do not show a strong correlation with
the degree of isolation of the 2D magnetic layers.

The analysis of magnetic ordering in zero applied field in terms of
inter-layer coupling $J_{\perp}$ presented here does not take into
account the effect of single-ion--type anisotropy on the magnetic
order. This has been suggested to be important close to $T_{\mathrm{N}}$
in several examples of 2D molecular magnet~\cite{Xiao2009-XY-q2DHA}
where it causes a crossover to $XY$-like behavior. In fact, its influence
is confirmed in the nonmonotonic $B$--$T$ phase diagram seen in
{[}Cu(HF$_{\text{2}}$)(pyz)$_{\text{2}}${]}BF$_{\text{4}}$. It
is likely that this is one factor that determines the ordering temperature
of a system, although, as shown in Ref.~\onlinecite{Xiao2009-XY-q2DHA},
it is a smaller effect than the interlayer coupling parametrized by
$J_{\perp}$. The future synthesis of single crystal samples of these
materials will allow the measurement of the single-ion anisotropies
for the materials studied here.

The presence of muon--fluorine dipole--dipole oscillations allows
the determination of some muon sites in these materials, although
it appears from our results that these are not those that lead to
magnetic oscillations. However, the \Fmu\  signal has been shown
to be useful in identifying transitions at temperatures well above
the magnetic ordering transition, which appear to have a structural
origin. The fluorine oscillations hamper the study of dynamic fluctuations
above $T_{\mathrm{N}}$, which often appear as a residual relaxation
on top of the dominant nuclear relaxation. It may be possible in future
to use RF radiation to decouple the influence of the fluorine from
the muon ensemble to allow muons to probe the dynamics.

The muon-spin precession signal, upon which much of the analysis presented
here is based, is seen most strongly in the materials containing Cu$^{\text{2+}}$
and is more heavily relaxed in the Ni$^{\text{2+}}$ materials. This
is likely due to the larger spin value in the Ni-containing materials.
This is borne out by measurements on pyz-based materials containing
Mn and Fe ions~\cite{Lancaster2006-Mndca2pyz-and-FeNCS2pyz2}, where
no oscillations are observed, despite the presence of magnetic order
shown unambiguously by other techniques. In the case of Mn-containing
materials magnetic order is found with \muSR\  through a change in
relative amplitudes of relaxing signals due to a differerence in the
nature of the relaxation on either side of the transition. It is likely,
therefore that muon studies of molecular magnetic materials containing
ions with small spin quantum numbers will be most fruitful in the
future.

Finally, the temperature dependence of the relaxation rates in these
materials has been shown to be quite complex, reflecting the variety
of muon sites in these systems. In favourable cases these data could
be used to probe critical behavior, such as critical slowing down,
although the unambiguous identification of such behavior may be problematic.

Despite these limitations on the use of \muSR\  in examining molecular
magnetic systems of the type studied here, it is worth stressing that
the technique still appears uniquely powerful in providing insights
into the magnetic behavior of these materials and will certainly be
useful in the future as a wealth of new systems are synthesised and
the goal of microscopically engineering such materials is approached.

%For the {[}Cu(HF$_{\text{2}}$)(pyz)$_{\text{2}}${]}\emph{X}
%system, tetrahedral
%anions \emph{X} give rise to significantly suppressed transition temperatures compared to
%similar materials with octahedral anions
%due to increased isolation of the layers. 
%The pyo-containing
%systems {[}Cu(pyz)$_{\text{2}}$(pyo)$_{\text{2}}${]}\emph{X}$_{\text{2}}$
%and {[}Cu(pyo)$_{\text{6}}${]}\emph{X}$_{\text{2}}$
%shows a weak anisotropy induced by Jahn--Teller distortions in the
%oxygen octahedra surrounding the copper ions. The two members
%of the {[}Ni(HF$_{\text{2}}$)(pyz)$_{\text{2}}${]}\emph{X} family
% contain $S=1$ nickel ions with octahedral anions,
%and a significant variation was found in $T_{\mathrm{N}}$ between
%them.

\begin{acknowledgments}
This work was partly supported by the Engineering and Physical Sciences
Research Council, UK. Experiments at the ISIS Pulsed Neutron and Muon
Source were supported by a beamtime allocation from the Science and
Technology Facilities Council. Further experiments were performed
at the Swiss Muon Source, Paul Scherrer Institute, Villigen, Switzerland.
This research project has been supported by the European Commission
under the 7$^{\text{th}}$ Framework Programme through the `Research
Infrastructures' action of the `Capacities' Programme, Contract No:
CP-CSA\_INFRA-2008-1.1.1 Number 226507-NMI3. The work at EWU was supported
by the National Science Foundation under grant no. DMR-1005825. Work
supported by UChicago Argonne, LLC, Operator of Argonne National Laboratory
(`Argonne'). Argonne, a US Department of Energy Office of Science
laboratory, is operated under Contract No. DE-AC02-06CH11357. The
authors would like to thank Paul Goddard, Ross McDonald, William Hayes
and Johannes Möller for useful discussions.
\end{acknowledgments}
\bibliographystyle{apsrev4-1}
\bibliography{2DMM}

\end{document}